\documentclass[a4paper, oneside, fleqn, 12pt]{scrartcl}

\usepackage[ngerman,USenglish]{babel}
\usepackage[utf8]{inputenc}
\usepackage{natbib}
\usepackage{amsmath,amssymb,bm}
\usepackage{textcomp}
\usepackage{dsfont}
\usepackage{color}
\usepackage{graphicx}
\usepackage{graphics}
\usepackage{multirow}
\usepackage{lscape}
\usepackage{tabularx}
\usepackage{tabulary}
\usepackage{booktabs}
\usepackage{authblk}
\usepackage{subfigure}
\usepackage{breakurl}
\usepackage[font={small,sl},labelfont=bf]{caption} 

\usepackage[running]{lineno}
\usepackage[pagebackref]{hyperref}
\hypersetup{%
	colorlinks=true,
	linkcolor=blue, 
	citecolor=blue,
	bookmarksopen=true,
	pdfstartpage={1},
	pdftitle={Spatial prediction of diameter distributions for the alpine protection forests in Ebensee, Austria, using ALS/PLS and spatial distributional regression models},
	pdfauthor={Arne Nothdurft, Andreas Tockner, Sarah Witzmann, Christoph Gollob, Tim Ritter, Ralf Kraßnitzer, Karl Stampfer, Andrew O. Finley},
	pdfkeywords={}
}

\newcommand{\bP}{ {\boldsymbol P} }

\newcommand{\bs}{ {\boldsymbol s} }

\newcommand{\bx}{ {\boldsymbol x} }

\newcommand{\by}{ {\boldsymbol y} }

\newcommand{\bbeta}{ {\boldsymbol \beta} }

\newcommand{\bnu}{ {\boldsymbol \nu} }

\newcolumntype{L}[1]{>{\raggedright\arraybackslash}p{#1}} 
\newcolumntype{C}[1]{>{\centering\arraybackslash}p{#1}} 
\newcolumntype{R}[1]{>{\raggedleft\arraybackslash}p{#1}} 

\newcommand{\beginsupplement}{%
        \setcounter{table}{0}
        \renewcommand{\thetable}{S\arabic{table}}%
        \setcounter{figure}{0}
        \renewcommand{\thefigure}{S\arabic{figure}}%
}

\graphicspath{{figures/}}

\title{\Large 
Spatial prediction of diameter distributions for the alpine protection forests in Ebensee, Austria, using ALS/PLS and spatial distributional regression models
}

\author[1]{\large Arne Nothdurft\thanks{\href{mailto:arne.nothdurft@boku.ac.at}{arne.nothdurft@boku.ac.at}, Tel: +43-1-47654-91411}}
\author[1]{\large Andreas Tockner}
\author[1]{\large Sarah Witzmann}
\author[1]{\large Christoph Gollob}
\author[1]{\large Tim Ritter}
\author[1]{\large Ralf Kraßnitzer} 
\author[2]{\large Karl Stampfer} 
\author[3]{\large Andrew O. Finley}

\affil[1]{\normalsize University of Natural Resources and Life Sciences, Vienna (BOKU), Department of Forest and Soil Sciences, Institute of Forest Growth, Austria}

\affil[2]{\normalsize University of Natural Resources and Life Sciences, Vienna (BOKU), Department of Forest and Soil Sciences, Institute of Forest Engineering, Austria}

\affil[3]{\normalsize Michigan State University, Departments of Forestry and Statistics \& Probability, USA}

\date{\large \today}

\begin{document}

\maketitle

\begin{abstract}
\noindent A spatial distributional regression model is presented to predict the forest structural diversity in terms of the distributions of the stem diameter at breast height (DBH) in the protection forests in Ebensee, Austria. In total 36,338 sample trees were measured via a handheld mobile personal laser scanning system (PLS) on 273 sample plots each having a 20\,m radius. Recent airborne laser scanning (ALS) data was used to derive regression covariates from the normalized digital vegetation height model (DVHM) and the digital terrain model (DTM). Candidate models were constructed that differed in their linear predictors of the two gamma distribution parameters. Non-linear smoothing splines outperformed linear parametric slope coefficients, and the best implementation of spatial structured effects was achieved by a Gaussian process smooth. 
Model fitting and posterior parameter inference was achieved by using full Bayesian methodology and MCMC sampling algorithms implemented in the \verb|R|-package \verb|BAMLSS|. Spatial predictions of stem count proportions per DBH classes revealed that regeneration of smaller trees was lacking in certain areas of the protection forest landscape. 

\vspace{0.25cm}

\end{abstract}

\section{Introduction}\label{sec:intro}


The forests of the Alps provide a wide range of ecosystem services. In addition to offering wood production and unique habitat for a diverse set of species, these forests protect people, buildings, and infrastructure from natural hazards such as snow avalanches, rockfalls, and mudslides \citep{Brang2001107}. Forests that protect human infrastructure are declared as ``protection forests,'' often through a public authority's formal decree. In Austria, between 2009 and 2015, there were 20--70 severe rockfall events per year, resulting in 10 human injuries and damage to settlement areas and infrastructure  \citep{Perzl2017}. Given its mountains terrain, $\sim$15.7\% (615,852 hectares, ha) of Austria's total forest area offer some level of protective function \citep{BML2022}.

In an in-situ rockfall experiment, \cite{Dorren2005183} show forest cover significantly reduced velocity, rebound height, residual hazard of rockfall, and depending on the quality and quantity of the forest structure, the number of rocks involved in a rockfall could be reduced by 64\%. To a large degree, protective forest effectiveness is a function of the forest's vertical and horizontal structure. Specifically, relevant structural measures are stem density, tree sizes (i.e., diameter), and patch size (see, e.g., \citealt{Brang2001107} and other references herein). To assess hazard risk and support the decision-making in forest management activities, \cite{Teich20091910} use computer simulation models informed  with spatially explicit forest structural summary inputs. These forest structure inputs were traditionally derived from aerial image analysis \citep{Bebi20013}. More recently, however, laser imaging detection and ranging (LiDAR) data are being used to support forest inventory \citep{Koehl2006}, and offer improved accuracy and efficiency through automation for mapping forest structure \citep{ADNAN2019111}. 

Mean tree diameter and total stem count are often inadequate at characterizing forest structure, especially for structurally diverse settings. Building structural diversity, through silvicultural treatments, is a common management objective that has been shown to enhance biological diversity, carbon storage, and possibly climate resilience \citep{Schuetz2002, DAmato2011, Gough2019}. This desired structural diversity is often produced by ``plentering,'' a highly intensive silvicultural prescription designed to move homogeneous stands to an uneven age structure with stratified crown layering \citep{Schuetz2001}. Characterizing such structurally diverse forests is best accomplished using more detailed summaries of possibly complex size-class distributions. 

Characterizing size-class distributions has traditionally been done using either a parameter prediction model (PPM) or a parameter recovery model (PRM) approach \citep{HyinkMoser1985}. In PPM, a probability density function (pdf) is chosen to characterize the size-distribution (e.g., diameter distribution), and the pdf parameters are then estimated separately for each sample plot. Finally, the parameter estimates are regressed against covariates via regression analysis. In PRM, the mean and dispersion parameters are directly regressed to meaningful characteristics, and the estimates of the pdf parameters are finally achieved by the ``method of moments.'' The PRM is often used to avoid confounding problems, which can occur with PPM, as similar pdfs can be achieved with different parameter combinations making it difficult to find unambiguously meaningful covariates. As consequence, the PPM equations can usually explain only little variation in the parameters and have typically a low R\textsuperscript{2}. A shortcoming of the traditional PPM and PRM approaches to model tree size distributions is that they require separate model steps---estimating size distribution parameters happens separate from regression used to explain variability in those parameters. This can easily produce ill-behaved prognoses of diameter distributions when new covariate data is used. To make future predictions of diameter distributions more reliable and accommodate temporal dependence among repeat measurements, a bivariate distribution modeling was demonstrated in \cite{KnoebelBurkhart1991}. \cite{Finley2014} proposed a non-parametric Bayesian approach to estimating diameter distributions that modeled each diameter class using a Poisson regression informed using LiDAR covariates and random effects designed to accommodate correlation among diameter classes and across spatial locations. While highly flexible, their proposed approach was computationally demanding and required a greater degree of user input to choose appropriate prior distributions and assess model convergence. 

Here, we demonstrate and assess a different inferential approach aimed at overcoming key limitations of previously proposed methods for characterizing size-class distributions. Specifically, we apply recent advancements in \emph{distributional regression} using generalized additive models that facilitate joint estimation of shape and scale parameters in parametric distributions. In particular we use a \emph{generalized additive models for location, scale, and shape }(GAMLSS) based approach proposed by \cite{RigbyStasinopoulos2005}. In a series of papers, \cite{Klein2015AOAS,Klein2015JASA,Klein2015JRSS} and \cite{Umlauf2018} extended the original maximum likelihood mode of inference for GAMLSS parameters to a Bayesian approach using Markov chain Monte Carlo (MCMC) referred to \emph{Bayesian additive models for location, scale, and shape }(BAMLSS). This Bayesian approach accommodate a richer set of models and uncertainty quantification. Many proposed BAMLSS features have been made available in user-friendly software \citep{Umlauf2021}.

Whereas within classical regression models, where the conditional mean of the response is regressed against covariates, distributional regression addresses the complete conditional response distribution, in that each distribution parameter is modeled in terms of covariates and, potentially, random effects. Compared to maximum likelihood based GAMLSS, the BAMLSS distributional regression supports a wider selection of distribution families, for which the parameters are not necessarily directly related to the location, scale, and shape of the given distribution but can form these measures indirectly via functional relationships. \cite{Kneib2023} offers an excellent review of distributional regression approaches including GAMLSS (and its Bayesian extensions) and the traditionally more conspicuous quantile regression. The review underscores key advantages to GAMLSS approaches with regard to modeling distribution parameters using versatile additive structure, nonlinear functions, varying coefficients, and spatially and temporally structured random effects.

In this paper, GAMLSS spatial distributional regression models are used to quantify forest structural diversity based on stand-level DBH distributions in a protection forest landscape near Ebensee, Austria. Sample plot data was collected using a handheld mobile personal laser scanning (PLS) and processed using automated software routines. Regression covariates were derived from a digital vegetation-height model (DVHM) and a digital terrain model (DTM) provided by recent airborne laser scanning (ALS) campaigns.

\section{Materials and methods}\label{sec:methods}

\subsection{Study region and model data}\label{sec:data}

\begin{figure*}[ht!]
\centering
\includegraphics[width=1.0\textwidth]{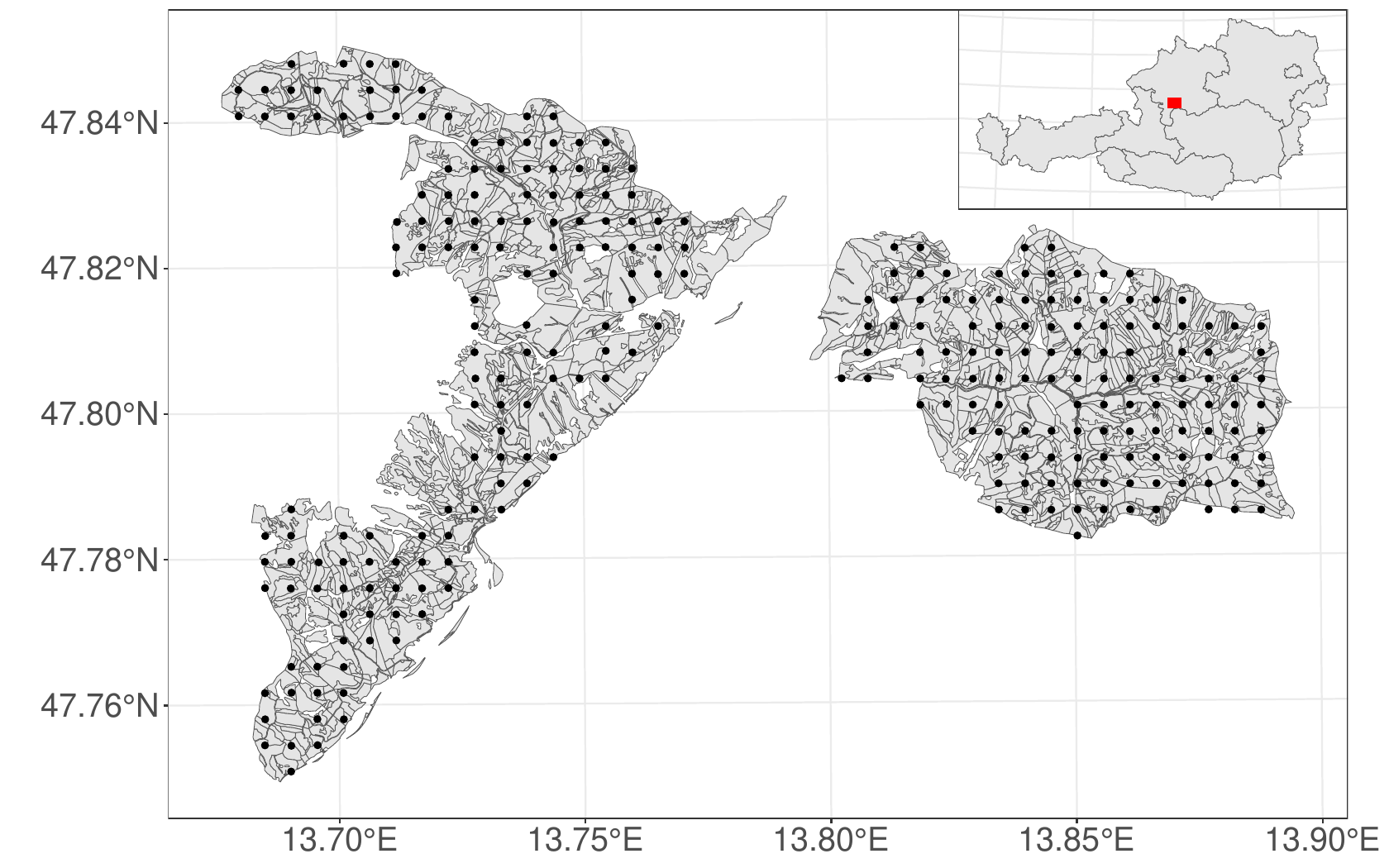}
\caption{Location and extent of the Austrian forest district Ebensee and locations of the 273 sample points in the southern region of the federal state of Upper Austria.}
\label{fig:map_sample_plots}
\end{figure*}

The study area was located in the southern region of the federal state of Upper Austria, near the village of Ebensee, and covers an area south of Traunsee lake (Fig.~\ref{fig:map_sample_plots}). The forest district Ebensee had a total area of 4,898 hectares (ha) and was partitioned into $Q=$1,237 forest stands. The average stand size was 3.96\,ha, the minimum 0.14\,ha, the median 2.33\,ha, and the maximum 89.99\,ha.

Forest inventory data was collected on $n=\text{273}$ sample plots, which were spatially aligned in a regular 400\,m$\times$400\,m grid (Fig.\,\ref{fig:map_sample_plots}). Plot measurements were collected using a handheld mobile PLS GeoSLAM ZEB Horizon (GeoSLAM Ltd., Nottingham, UK). The 180 plots in the study area's Eastern half were scanned in autumn 2021, and the remaining 93 plots in the Western half were scanned in spring 2023. Position, diameter at breast height (DBH), and height for the approximately 36,338 measurement trees were derived from 3D point clouds collected on each 20\,m radius plot centered on the $n$ grid locations using fully automated routines detailed in \cite{Gollob2019,Gollob2020}, in \cite{Tockner2022} and in \cite{Ritter2017,Ritter2020}. Stem volume was calculated using a traditional stem-form function \citep{Pollanschuetz1965}. The mean DBH of measurement trees was 14.6\,cm, the standard deviation was 10.6, the minimum was equal to the pre-defined 5\,cm threshold, and the maximum DBH was 92.4\,cm. For each plot, growing stock timber volume (GSV) was expressed as m$^3$/ha (i.e., computed as the sum of tree volume scaled by the 7.958 fixed-area plot tree expansion factor). The mean GSV of the sample plots was 259.6\,m$^3$/ha, the standard deviation was 177.9\,m$^3$/ha, and the minimum and maximum were 0.6\,m$^3$/ha and 980.7\,m$^3$/ha, respectively.

The federal state of Upper Austria provided open access to a DTM and a DSM via the open data platform \texttt{data.ooe.gv.at} \citep{LandOberosterreich2023b, LandOberosterreich2023a}; both were available as 0.5\,m$\times$0.5\,m resolution grids in the tagged image file format. The DTM and DSM were processed from ALS data obtained in different flight campaigns conducted over the past several years. For 84\% of the total forest district area, the ALS data were collected in 2021, for 11\% in 2019, and for the remaining 5\% in 2017. A normalized DVHM was computed by substracting the DTM from the DSM.

\subsection{Model construction}\label{sec:model}

A distributional regression model was built for the DBH distributions observed at the $n$ PLS forest inventory plots with the model form
\begin{linenomath*} 
\begin{equation}\label{eq:distribution_model}
 \by_i|\bx_i \sim \mathcal{D} \left(\vartheta_1(\bx_i),\ldots,\vartheta_K(\bx_i)\right), 
\end{equation}
\end{linenomath*} 
where $\by_i$ is the DBH distribution vector at the $i$-th plot, $\mathcal{D}$ is a  parametric density distribution function with parameters $\vartheta_1(\bx_i),\ldots,\vartheta_K(\bx_i)$ that depend on a set of plot specific covariates $\bx_i$.

The parameters are usually not directly formed by a regression predictor, but often through a monotonically increasing response function, which maps the predictor $\eta^{\vartheta_l}_i$ to the $l$-th parameter via
\begin{linenomath*} 
\begin{equation}\label{eq:response_function}
 \vartheta_{il} = h_l(\eta^{\vartheta_l}_i) \ .
\end{equation}
\end{linenomath*} 
Assuming monotony and inverting the response function via the link-function $g_l(\cdot)$ achieves the parameter predictor 
\begin{linenomath*} 
\begin{equation}\label{eq:link_function}
 \eta^{\vartheta_l}_i = h_l^{-1}(\vartheta_{il}) = g_l(\vartheta_{il})  \ .
\end{equation}
\end{linenomath*} 
The vector of covariates $\bx_i'=\left(\mathbf{x}_i',\bnu_i',\bs_i'\right)$ optionally contains measures $\mathbf{x}_i'$ having a linear effect, $\bnu_i'$ having nonlinear effects, and $\bs_i$ representing generic geo-locations.
Hence, the structured additive predictor becomes
\begin{linenomath*} 
\begin{equation}\label{eq:linear_parameter_predictor}
 \eta^{\vartheta_l}_i = \mathbf{x}_i'\bbeta^{\vartheta_l}+\sum_{j=1}^{J_l}f_j^{\vartheta_l}(\bnu_i)+f_{\text{geo}}^{\vartheta_l}(\bs_i) \ ,
\end{equation}
\end{linenomath*} 
and is composed of linear covariate effects with parameters $\bbeta^{\vartheta_l}$ in the first summand, smooth nonlinear functions $f_j^{\vartheta_l}(\cdot)$ in the second summand, and a spatial effect $f_{\text{geo}}^{\vartheta_l}(\cdot)$ at geo-locations $\bs_i$ in the third summand.

Distributional regression models were constructed with the gamma distribution as proper candidate for $\mathcal{D}$. Trials were also made with the Weibull distribution, but the Weibull distribution proved less flexible than the gamma distribution. Following \cite{Rigby2019}, the gamma distribution's probability density function (pdf) considered here is defined by
\begin{linenomath*} 
\begin{equation}\label{eq:gamma_pdf}
f_{\mathrm{GA}} (y| \mu, \sigma) = \frac{y^{1/\sigma^2-1}e^{-y/(\sigma^2\mu)}}{(\sigma^2\mu)^{1/\sigma^2}\Gamma(1/\sigma^2)} \ ,
\end{equation}
\end{linenomath*} 
for $y>0$, and with $\mu>0$ and $\sigma>0$. Herein, $\mathrm{E}(Y)=\mu$ and $\mathrm{Var}(Y)=\sigma^2\mu^2$. Linear predictors (Eq.\,\ref{eq:linear_parameter_predictor}) were constructed for both parameters $\mu$ and $\sigma$.

The catalogue of the possible covariates for $\mathbf{x}_i$ and $\bnu_i$ included summary statistics of the DVHM across the pixels per sample plot area, such as the mean vegetation height (MVH), its standard deviation (SDVH), and various percentiles of the distributions of the pixellated vegetation heights. In addition, topographic metrics were derived from the DTM, i.e., the elevation above sea level (ESL), and the average slope (SLO), and aspect (ASP) of the terrain. Finally, the geo-locations of the sample plot centroids were used to index a spatial Gaussian process (i.e., the $f_{\text{geo}}^{\vartheta_l}(\bs_i)$ summand in \ref{eq:linear_parameter_predictor}).

Various candidate models were tested that differed in their complexity, especially in terms of smooth nonlinear models for the covariate effects versus simple linear parametric coefficients, and with respect to presence/absence of a structured spatial effect.

The model fitting and subsequent prediction was performed within a Bayesian inferential framework and by using the \verb|R|-package \verb|BAMLSS| \citep{Umlauf2018, Umlauf2021}. Smooth nonlinear covariate effects (second summand in Eq.\,\ref{eq:linear_parameter_predictor}) were consistently modeled with \verb|BAMLSS|'s default thin plate regression splines \citep{Wood2003}, and the spatially structured effect for the continuously indexed sample plot location coordinates (third summand in Eq.\,\ref{eq:linear_parameter_predictor}) was alternatively represented by a spatial Gaussian process model or by a bivariate tensor product smooth. When a Gaussian process was chosen, a simplified form of the Matern covariance function was applied, according to suggestions by \cite{KammannWand2003}. Parameter inference was derived from posterior distributions that were sampled via MCMC techniques. Computations were performed on a multi-core processor workstation. On 7 cores, 5,000 iterations were computed per each core. From each of the 7 chains, the first 2,000 iterations were discarded as ``burn-in,'' and the from the remaining 3,000 iterations every 10-th sample was kept. This burn-in and thinning yielded approximately $M=$2,100 nearly independent MCMC samples upon which parameter and predictive posterior inference was based. The performances of the different candidate models were assessed by means of the deviance information criterion (DIC) \citep{Spiegelhalter2002} and two varieties of the widely applicable information criterion (WAIC1 and WAIC2) defined in \cite{Gelman2014}. For both, DIC and WAIC lower values indicate improved model fit.

\subsection{Prediction}\label{sec:prediction}

Our primary interest was in stem diameter distribution prediction for each of the 1,237 forest stands delineated in Fig.~\ref{fig:map_sample_plots}. For this purpose, the entire Ebensee forest district domain was partitioned into 35.5\,m$\times$35.5\,m squared prediction pixels, where each pixel's area equals that of the 20\,m radius sample plot. For each of the prediction pixels the same set of covariates as used in the candidate models were derived. Then, given these prediction pixel covariate values and centroid coordinates, posterior predictive distribution samples were generated via composition sampling for each pixel's $\mu$, $\sigma$, and subsequently gamma pdf based stem diameter distribution. 

To assess the protective function of the forest stands in terms of their structural diversity, forest practitioners from the Austrian Federal Forest Service were especially interested in the percentage shares of the stems that were allocated to broader diameter classes: (1) small (DBH$<$25\,cm), (2) intermediate (25\,cm$\leq$DBH$\leq$50\,cm), and (3) large (DBH$>$50\,cm).

To produce such estimates, the gamma distribution function $F_{GA}(\cdot)$ was evaluated with the $\mu^{(q)}_{j,m}$ and $\sigma^{(q)}_{j,m}$ estimates from each posterior sample $m=1,\ldots,M$ for each prediction pixel $j=1,\ldots,J_{q}$ of the in total $J_{q}$ pixels within each forest stand indexed by $q=1,\ldots,1237$ via: (1) $\bP^{(q)}_{j,m}(\text{DBH}< 20\,cm)=F_{GA}(y=20|\mu^{(q)}_{j,m},\sigma^{(q)}_{j})$, (2) $\bP^{(q)}_{j,m}(20\leq\text{DBH}\leq 45\,cm)=F_{GA}(y=45|\mu^{(q)}_{j},\sigma^{(q)}_{j})-F_{GA}(y=20|\mu^{(q)}_{j},\sigma^{(q)}_{j})$, and (3) $\bP^{(q)}_{j,m}(\text{DBH}> 45\,cm)= 1 - F_{GA}(y=45|\mu^{(q)}_{j},\sigma^{(q)}_{j})$.

Complete posterior predictive distributions of the $M$ aggregated estimates per forest stand were generated by an area-weighting through 
\begin{linenomath*} 
\begin{equation}\label{eq:dbh_class_stand_prediction}
\bP^{(q)}_{m}(\cdot)=\frac{1}{\sum_{j=1}^{J_q}a_j}\sum_{j=1}^{J_{q}}a_j\bP^{(q)}_{j,m}(\cdot) \ ,
\end{equation}
\end{linenomath*} 
with $a_j$ being the non-constant area of pixel $j$ that falls into stand $q$, and that might be reduced by stand border intersections.

\section{Results}\label{sec:results}

\subsection{Candidate models}\label{sec:candidate_models}

In total 15 candidate models were constructed that differed in their covariates and in their constructions of the linear predictors for the $\mu$ and $\sigma$ parameter of the gamma distribution (Table\,\ref{tab:candidate_models}). The covariate effects were either modeled through a linear trend that was represented by a single parametric slope coefficient, or via non-parametric smoothing splines. The effect of the terrain aspect (ASP) was throughout represented by a cyclic version of a cubic regression spline smooth.
The spatially structured effects were either modeled by a bivariate tensor product smooth with the continuous x,y-coordinates of the sample plots, or alternatively, by Gaussian process.

The distributional regression framework provided high flexibility and generally enabled different specifications of the linear predictors for both the $\mu$ and $\sigma$ parameter of a single distributional regression model. However, it was found that a unique specification of both linear predictors worked well throughout all candidate models. Consequently, the two linear predictors of the $\mu$ and $\sigma$ parameter of each distributional regression model were constructed with the same set of covariates and by using the same model representations (parametric term vs. smoothing spline) for the respective covariate effects.

Comparisons of the model performances in terms of the DIC and two calculations of the WAIC suggested that smoothing splines were more useful than the parametric linear trends; see diagnostics in Table~\ref{tab:candidate_models} for m\_2 versus (vs.) m\_1, m\_5 vs.~m\_4, m\_8 vs.~m\_6, m\_9 vs.~m\_7, m\_14 vs.~m\_12, and m\_15 vs.~m\_13.
Our findings also suggest that a spatially structured effect always enhanced the model performance, although this was generally associated with an increased number of effective model parameters (edf, p1, p2); compare m\_6 \& m\_7 vs.~m\_1, m\_8 \& m\_9 vs. m\_2, m\_10 \& m\_11 vs. m\_3, m\_12 \& m\_13 vs. m\_4, and m\_14 \& m\_15 vs. m\_5. When a spatially structured effect was considered, a Gaussian process proved more appropriate than a tensor product smooth; compare m\_6 vs.~m\_7, m\_8 vs.~m\_9, m\_10 vs.~m\_11, m\_12 vs.~m\_13, and m\_14 vs.~m\_15. Among all 15 candidate models, model m\_14 had a marginally lower DIC and WAIC and hence was considered as ``best'' and used for subsequent diameter distribution predictions.

\begin{landscape}

\begin{table}[ht!]
  \caption{Construction of the candidate models. ``p'' stands for a parametric effect and ``s()'' for a non-parametric smoothing spline. ``s(cc)'' indicates a cyclic version of a regression spline smooth. Spatially structured effects are either modeled by assuming a Gaussian process ``s(gp)'' or via a tensor product smooth ``te''. Diagnostics show the deviance information criterion (DIC), two calculation variants of the widely applicable information criterion (WAIC1, WAIC2), and the associated effective number of model parameters (edf, p1, p2). Lowest DIC and WAIC are in bold.}
  \label{tab:candidate_models}
  \begin{center}
    \resizebox{\columnwidth}{!}{\begin{tabular}{lccccccccccccccc}
    \toprule
       & m\_1 & m\_2 & m\_3 & m\_4 & m\_5 & m\_6 & m\_7 & m\_8 & m\_9 & m\_10 & m\_11 & m\_12 & m\_13 & m\_14 & m\_15\\ \midrule
MVH    & p & s & s     & s     & s     & p     & p    & s     & s    & s     & s     & s     & s     & s     & s     \\
SDVH   & p & s & s     & s     & s     & p     & p    & s     & s    & s     & s     & s     & s     & s     & s     \\
P2.5   &   &   &       & p     & s     &       &      &       &      &       &       & p     & p     & s     & s     \\
P97.5  &   &   &       & p     & s     &       &      &       &      &       &       & p     & p     & s     & s     \\
ESL    & p & s & s     & s     & s     & p     & p    & s     & s    & s     & s     & s     & s     & s     & s     \\
SLO    &   &   & s     & s     & s     &       &      &       &      & s     & s     & s     & s     & s     & s     \\
ASP    &   &   & s(cc) & s(cc) & s(cc) &       &      &       &      & s(cc) & s(cc) & s(cc) & s(cc) & s(cc) & s(cc) \\
XY     &   &   &       &       &       & s(gp) & te   & s(gp) & te   & s(gp) & te    & s(gp) & te    & s(gp) & te    \\
    \midrule
DIC    & 231.268 & 229.278 & 228.321 & 228.087 & 227.304 & 230.008 & 230.235 & 228.347 & 228.460 & 227.577 & 227.630 & 227.214 & 227.236 & \textbf{226.486} & 226.526 \\
edf    & 8.1 & 53.2 & 85.8 & 87.3 & 114.2 & 62.6 & 51.2 & 107.9 & 94.4 & 137.6 & 129.4 & 139.1 & 132.1 & 166.1 & 158.1  \\
WAIC1 & 231.269 & 229.277 & 228.319 & 228.086 & 227.302 & 230.005 & 230.233 & 228.343 & 228.458 & 227.574 & 227.627 & 227.211 & 227.234 & \textbf{226.484} & 226.524 \\
WAIC2 & 231.269 & 229.278 & 228.321 & 228.088 & 227.305 & 230.006 & 230.234 & 228.346 & 228.460 & 227.577 & 227.631 & 227.215 & 227.238 & \textbf{226.489} & 226.529 \\
p1 & 8.4 & 52.7 & 83.9 & 85.7 & 112.2 & 60.0 & 49.7 & 104.3 & 91.9 & 134.5 & 126.9 & 135.9 & 129.9 & 163.8 & 156.6 \\
p2 & 8.4 & 53.3 & 84.8 & 86.7 & 113.7 & 60.5 & 50.1 & 105.6 & 93.0 & 136.4 & 128.5 & 137.8 & 131.7 & 166.6 & 159.2 \\
    \bottomrule
  \end{tabular}}
  \end{center}
\end{table}

\end{landscape}

\subsection{Analysis and inference of the best model}\label{sec:model_analysis_inference}

\begin{figure*}
\begin{minipage}[c]{0.245\textwidth}\centering
\includegraphics[width=1\textwidth]{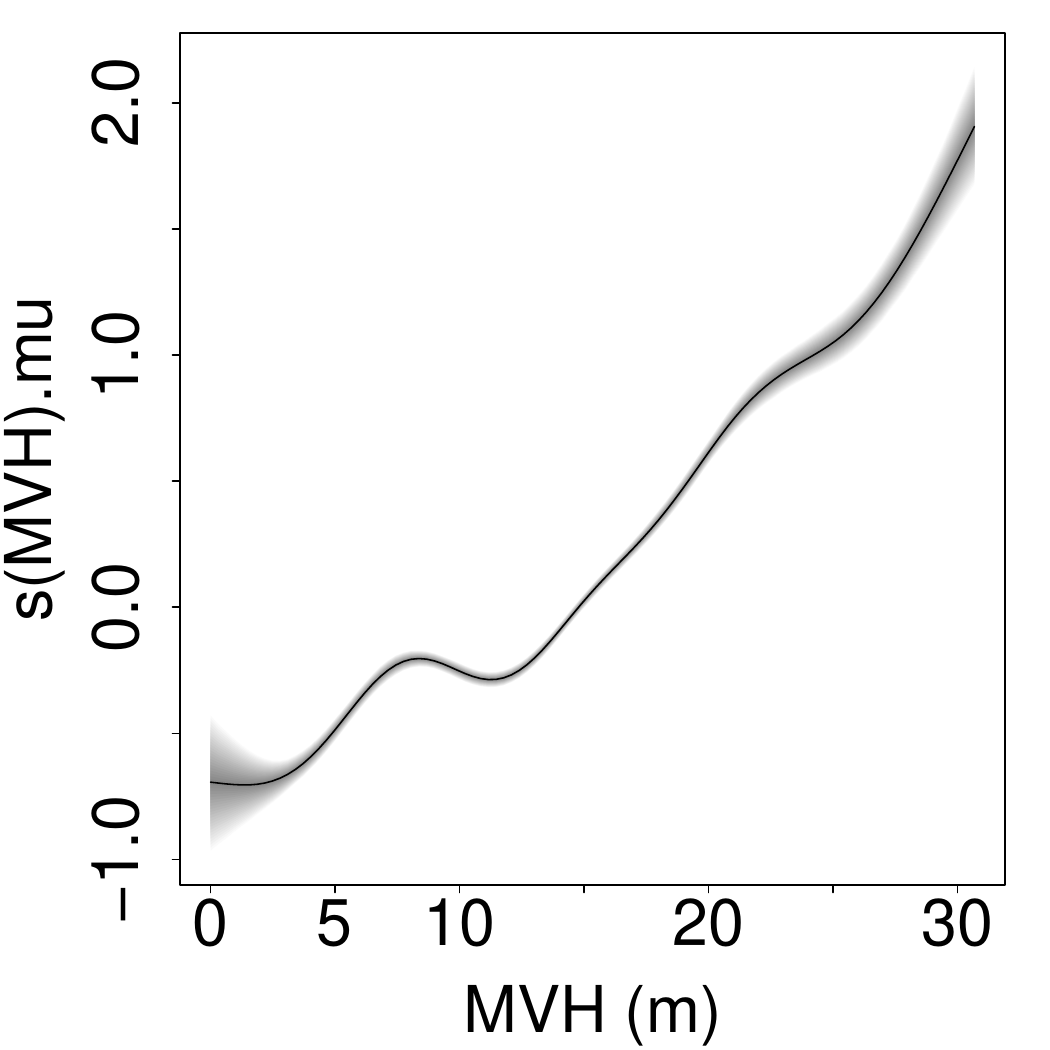}\\ 
\includegraphics[width=1\textwidth]{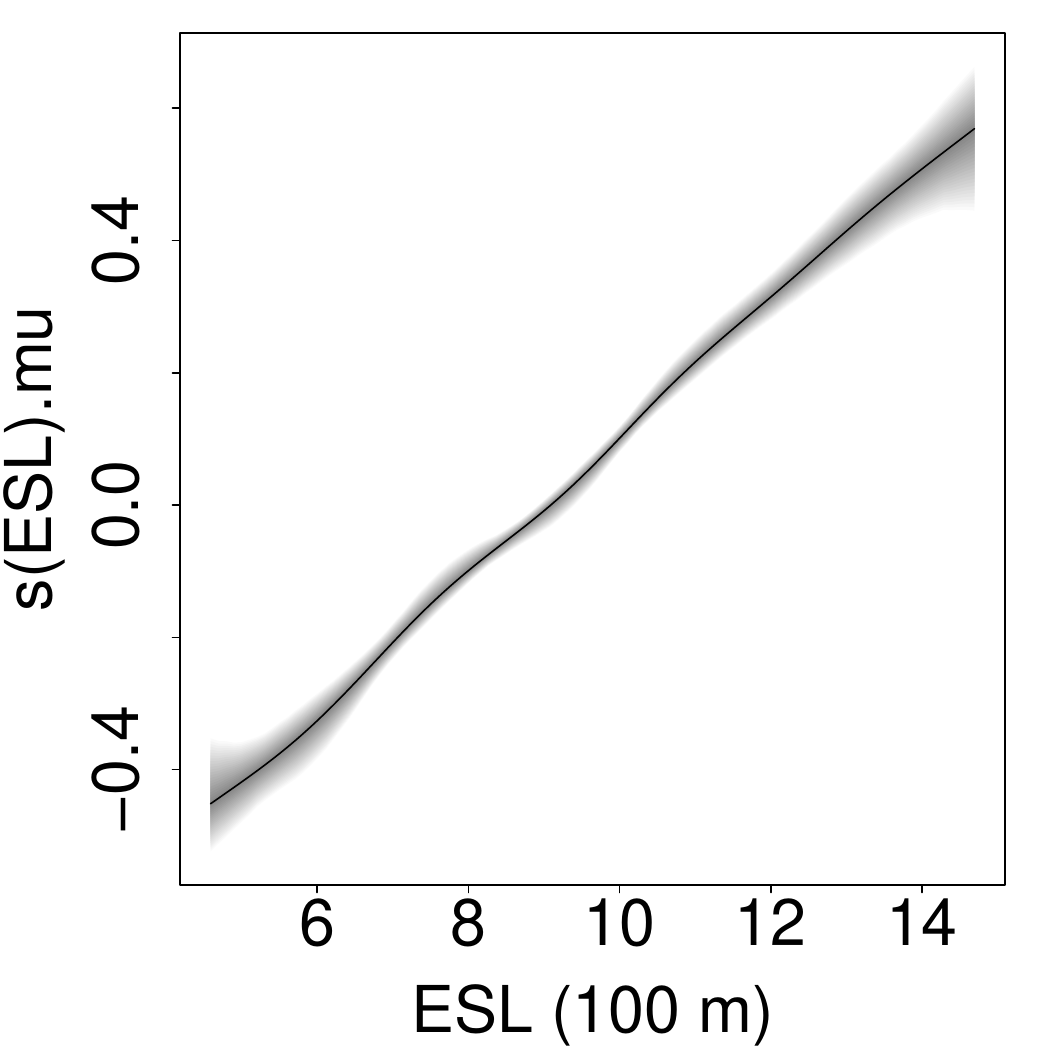}\\
\includegraphics[width=1\textwidth]{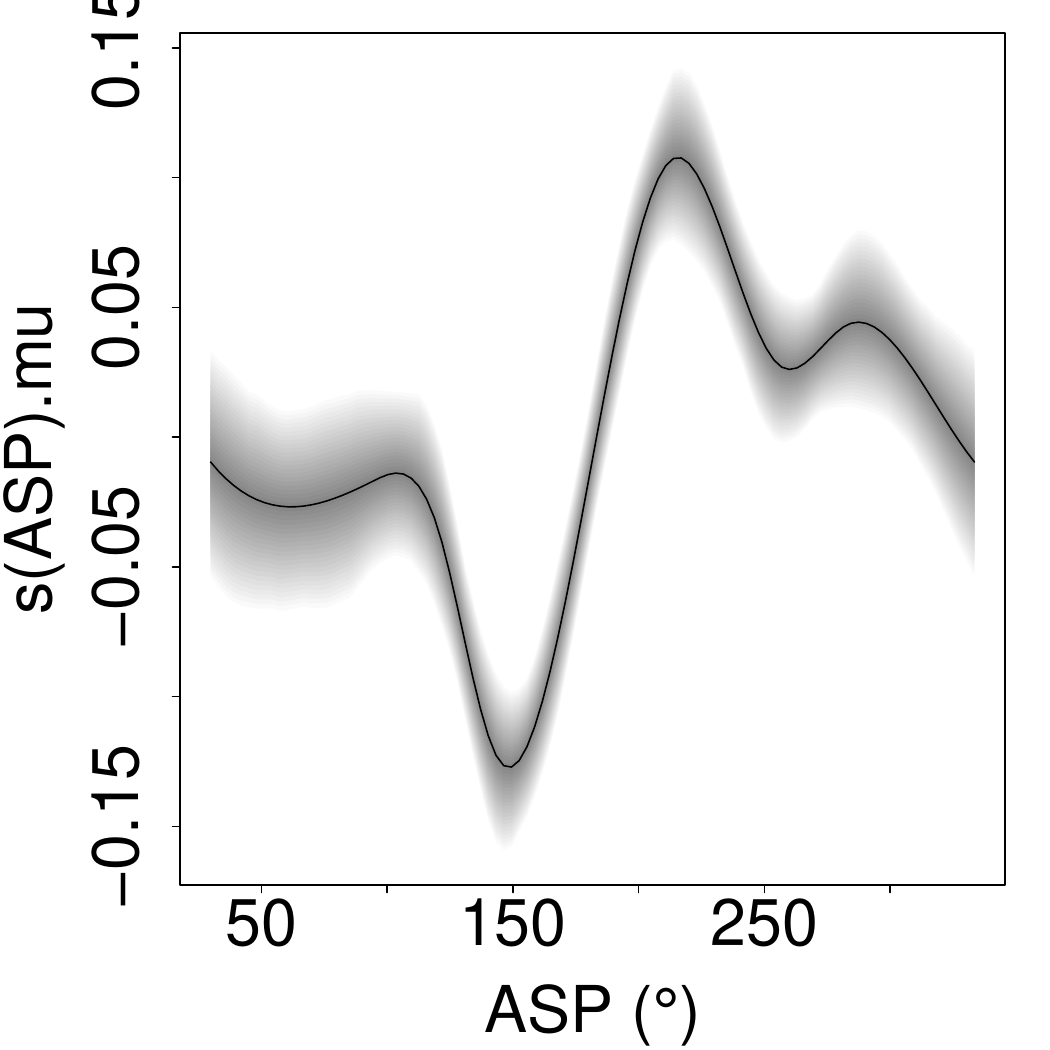}\\
\includegraphics[width=1\textwidth]{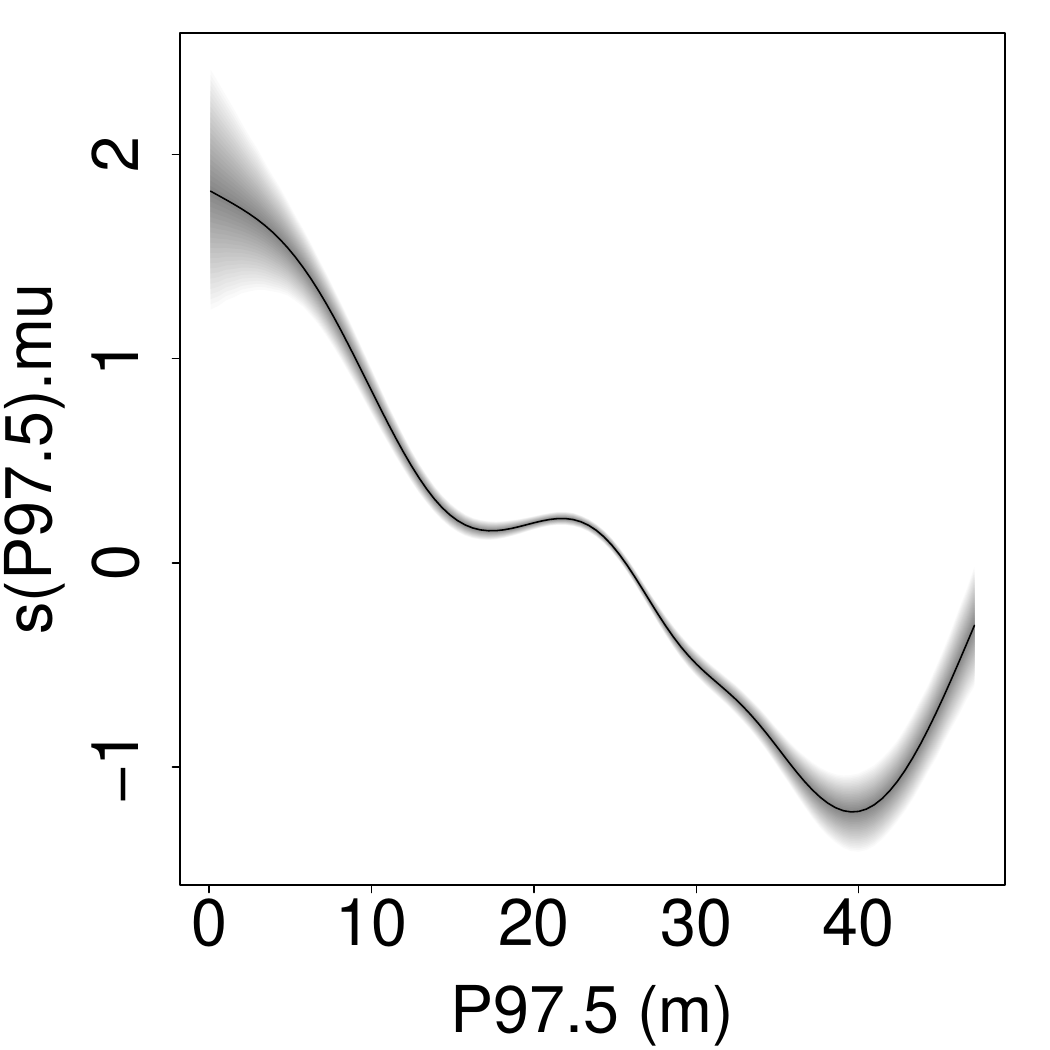}
\end{minipage}
\begin{minipage}[c]{0.245\textwidth}\centering
\includegraphics[width=1\textwidth]{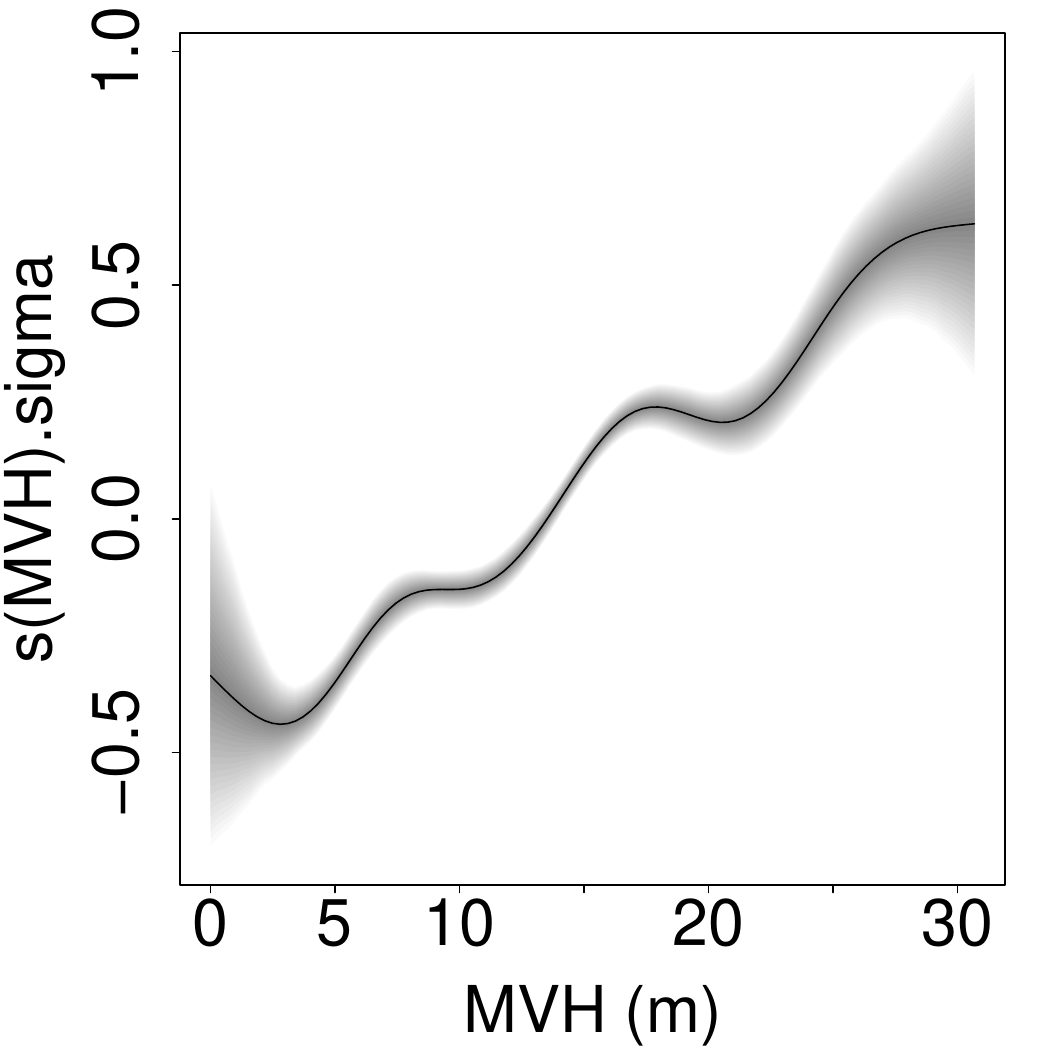}\\ 
\includegraphics[width=1\textwidth]{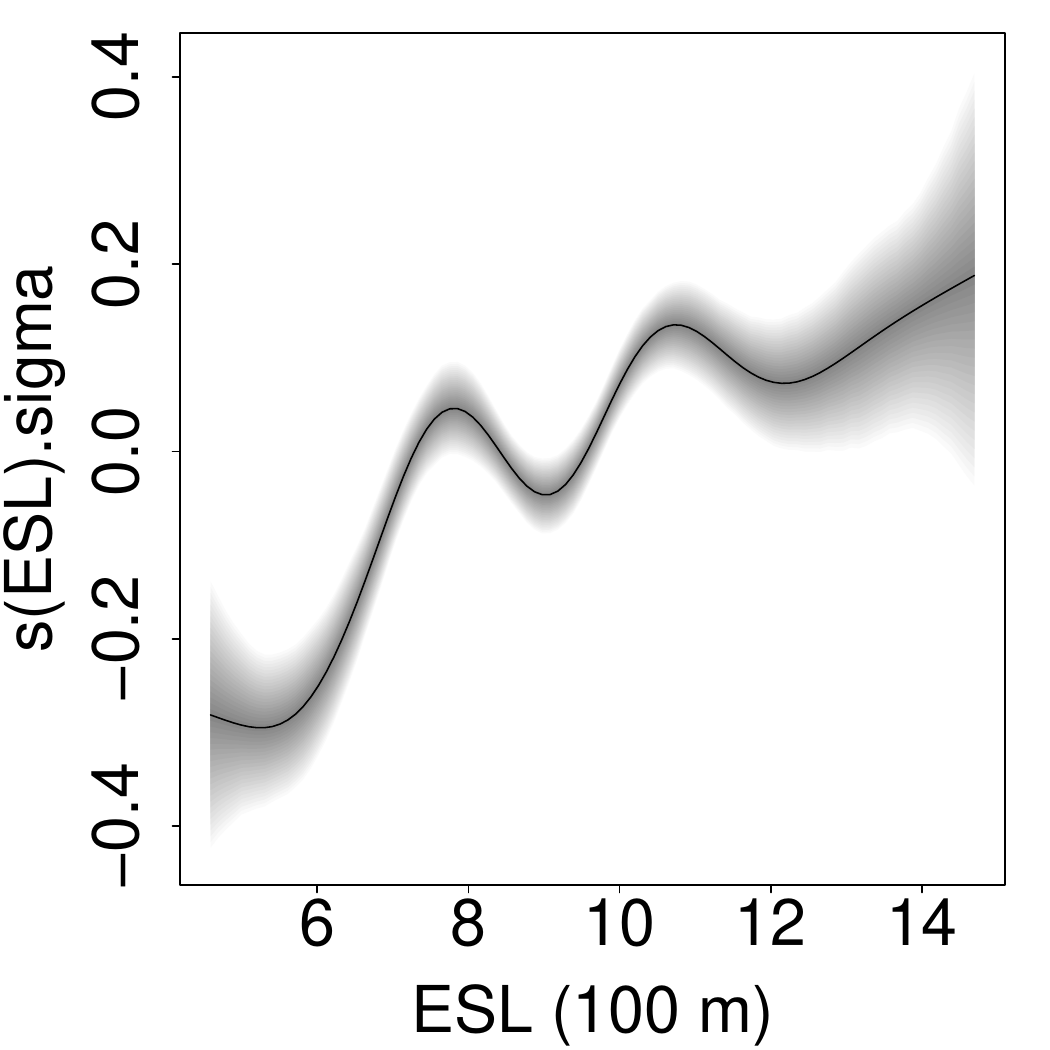}\\
\includegraphics[width=1\textwidth]{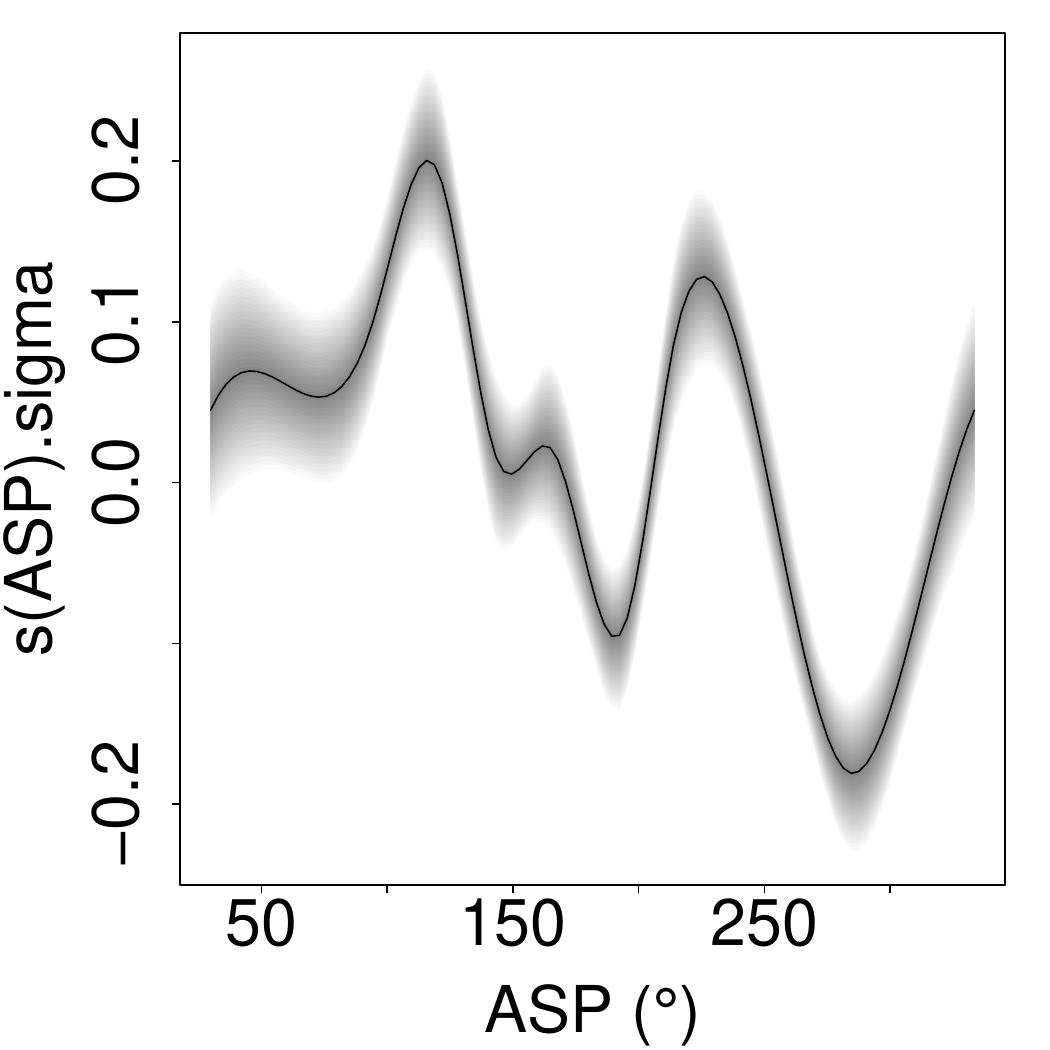}\\
\includegraphics[width=1\textwidth]{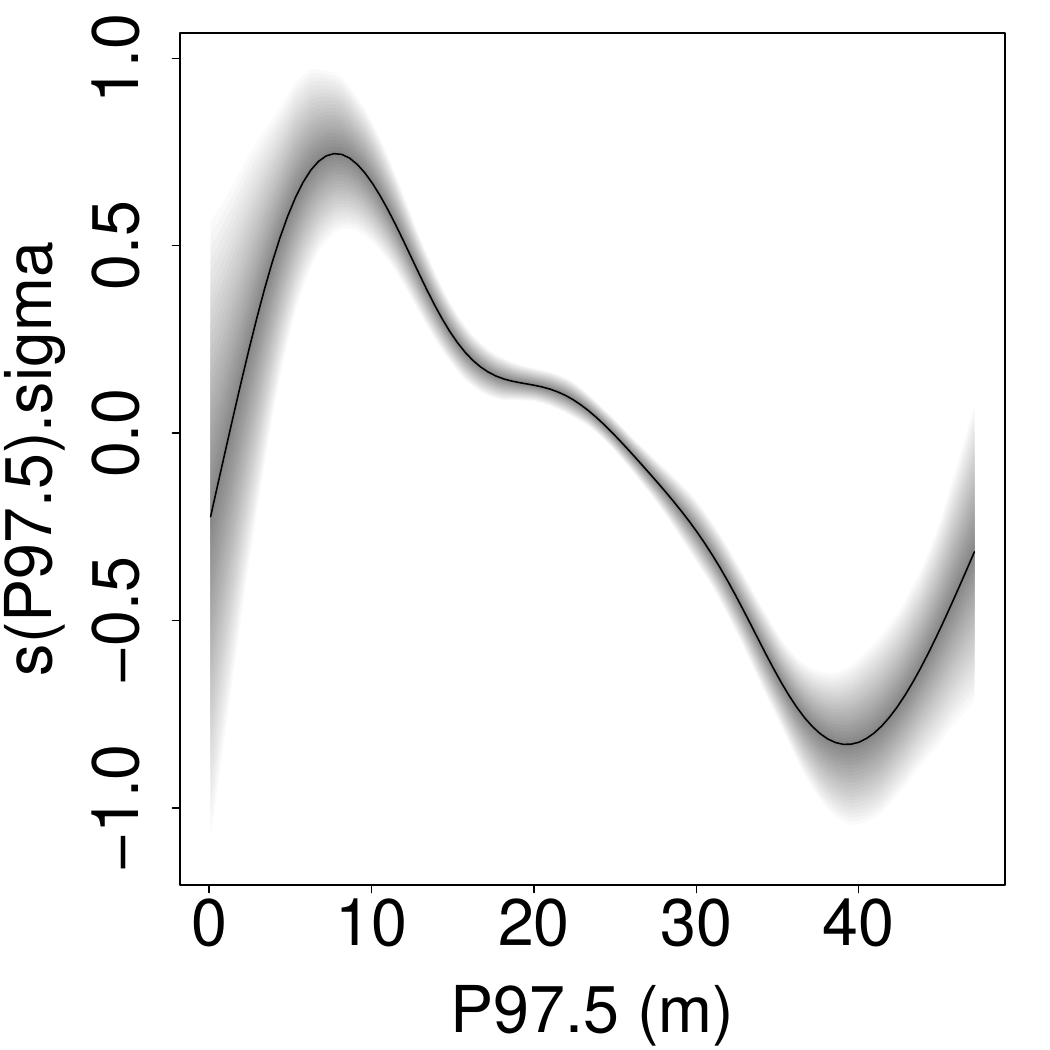}
\end{minipage}
\begin{minipage}[c]{0.245\textwidth}\centering
\includegraphics[width=1\textwidth]{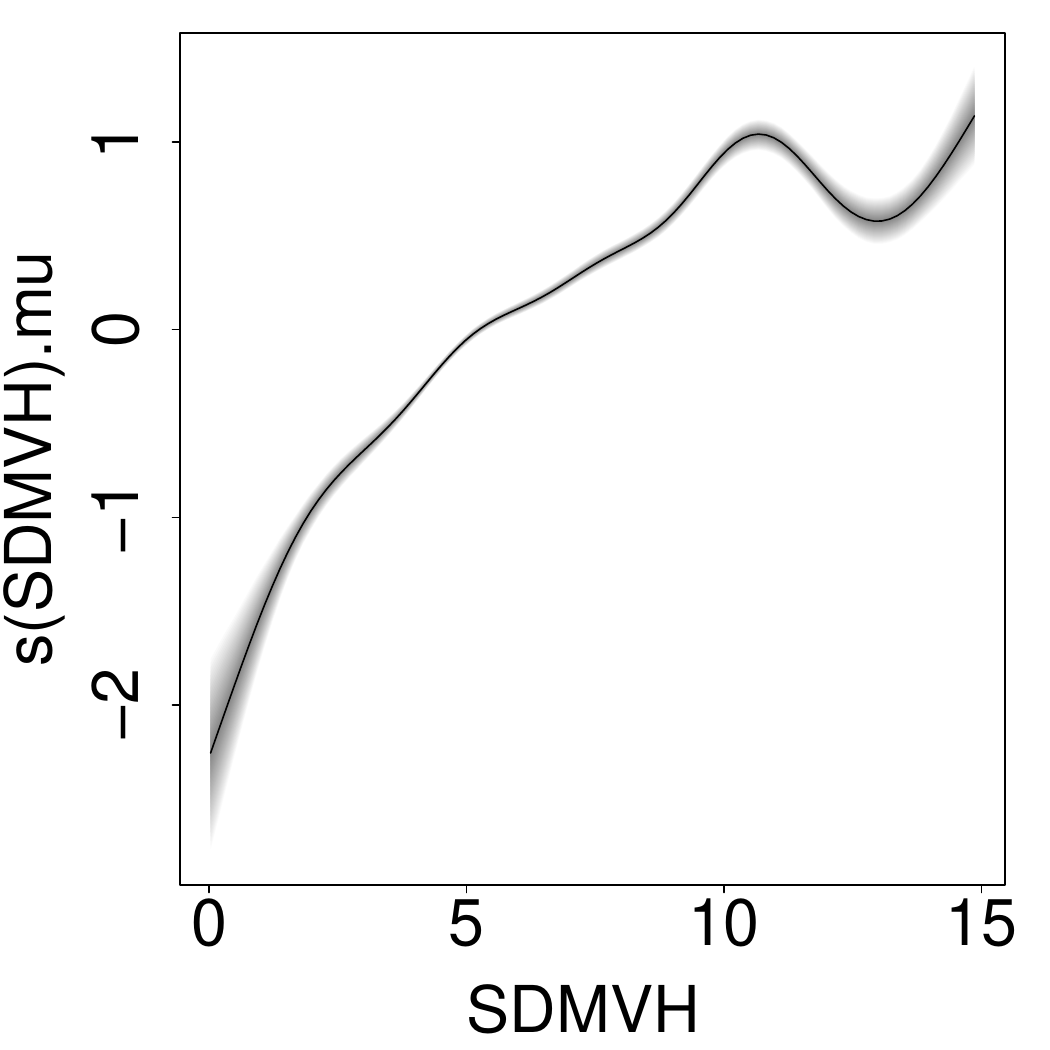}\\ 
\includegraphics[width=1\textwidth]{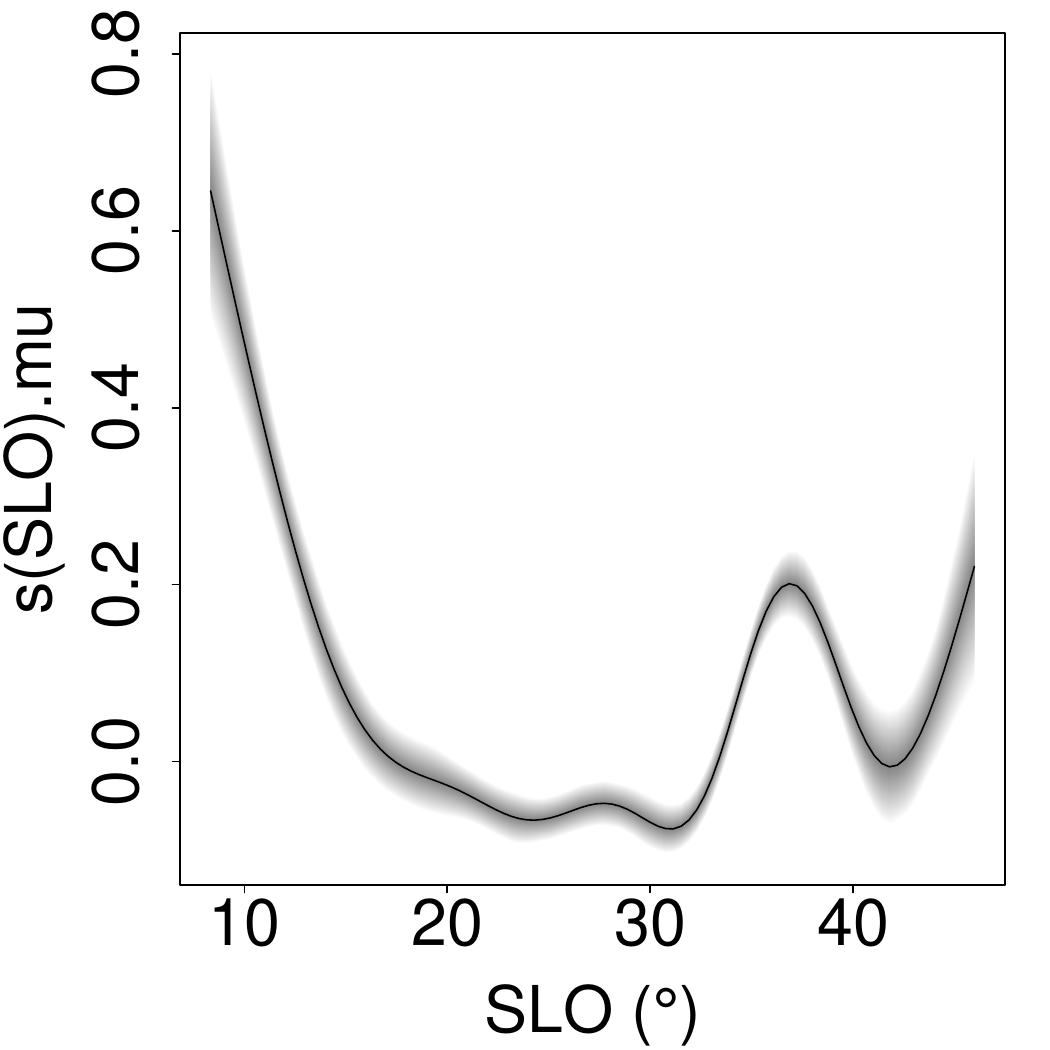}\\
\includegraphics[width=1\textwidth]{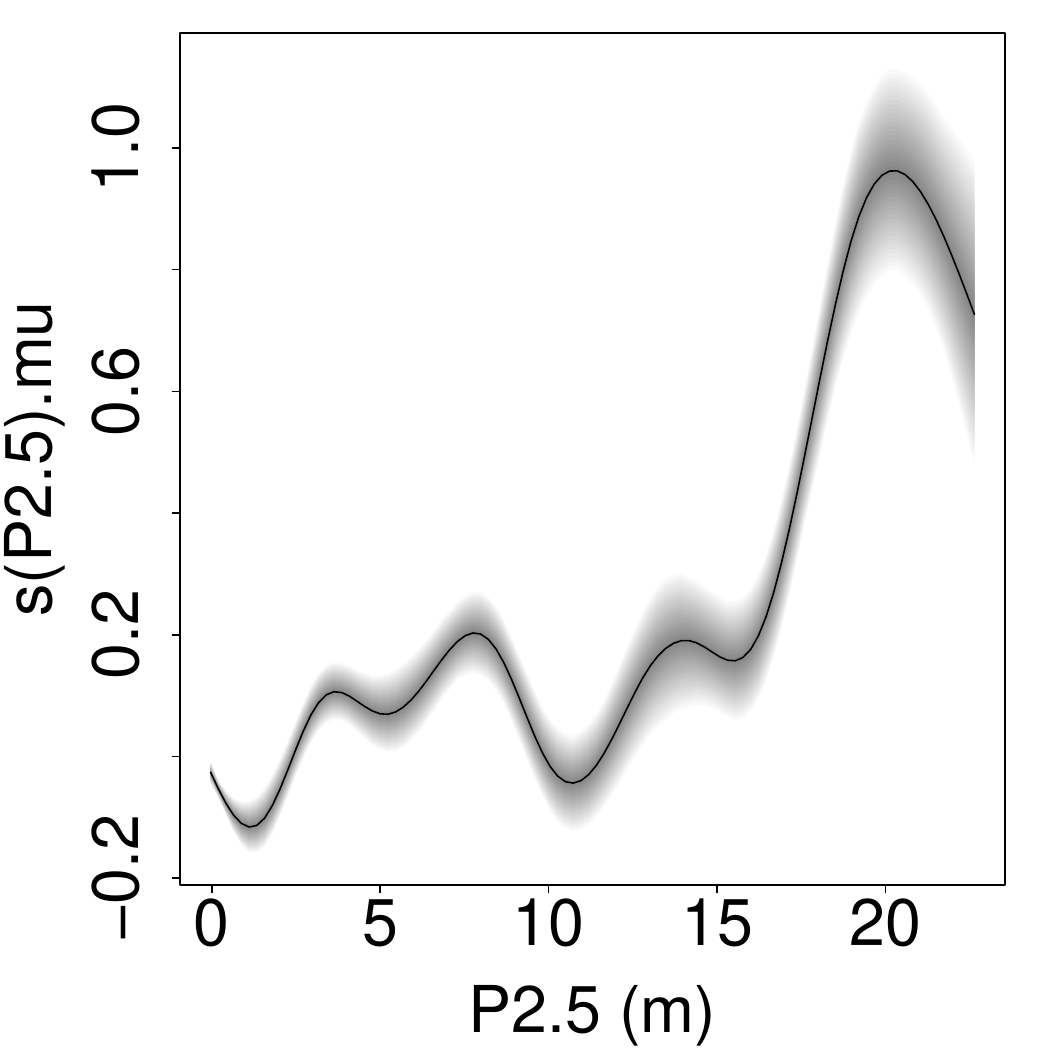}\\
\includegraphics[width=1\textwidth]{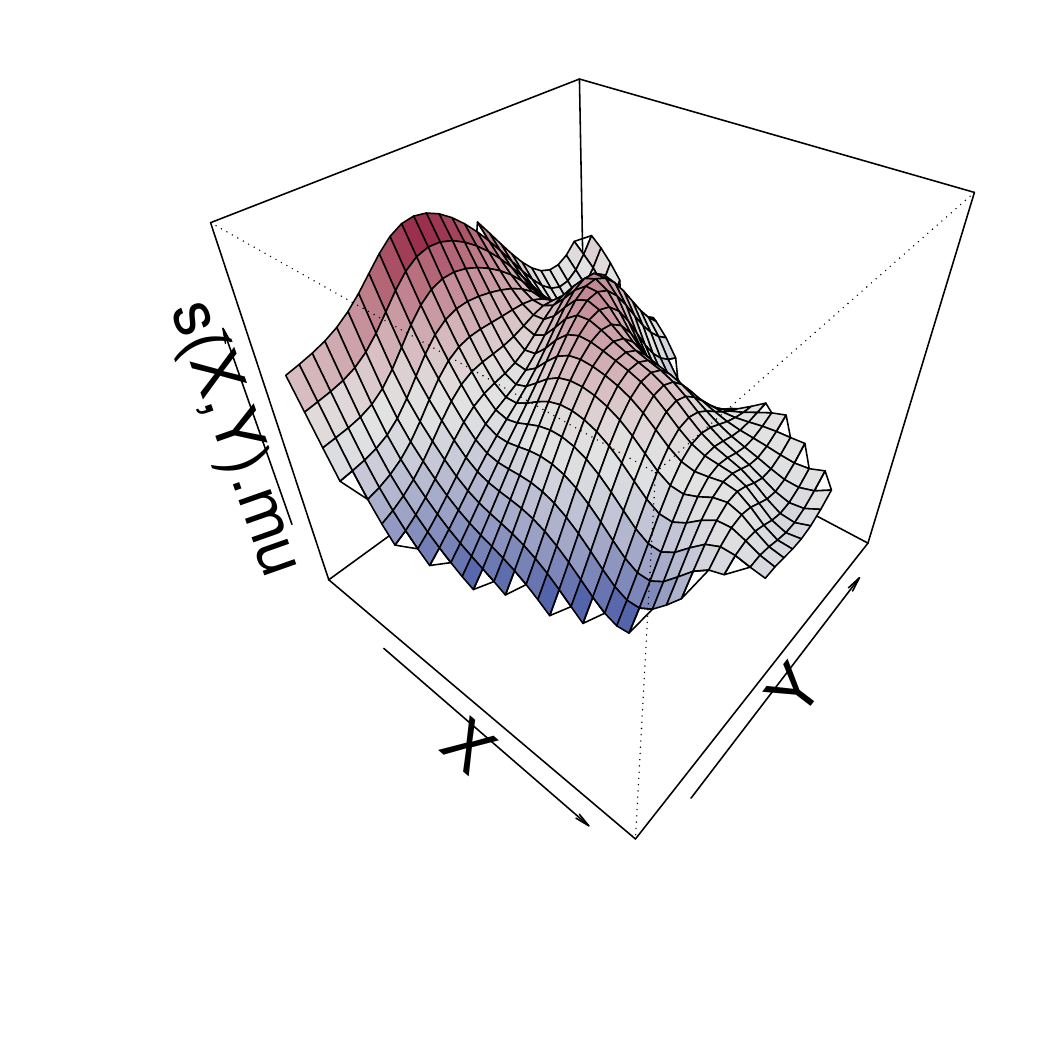}
\end{minipage}
\begin{minipage}[c]{0.245\textwidth}\centering
\includegraphics[width=1\textwidth]{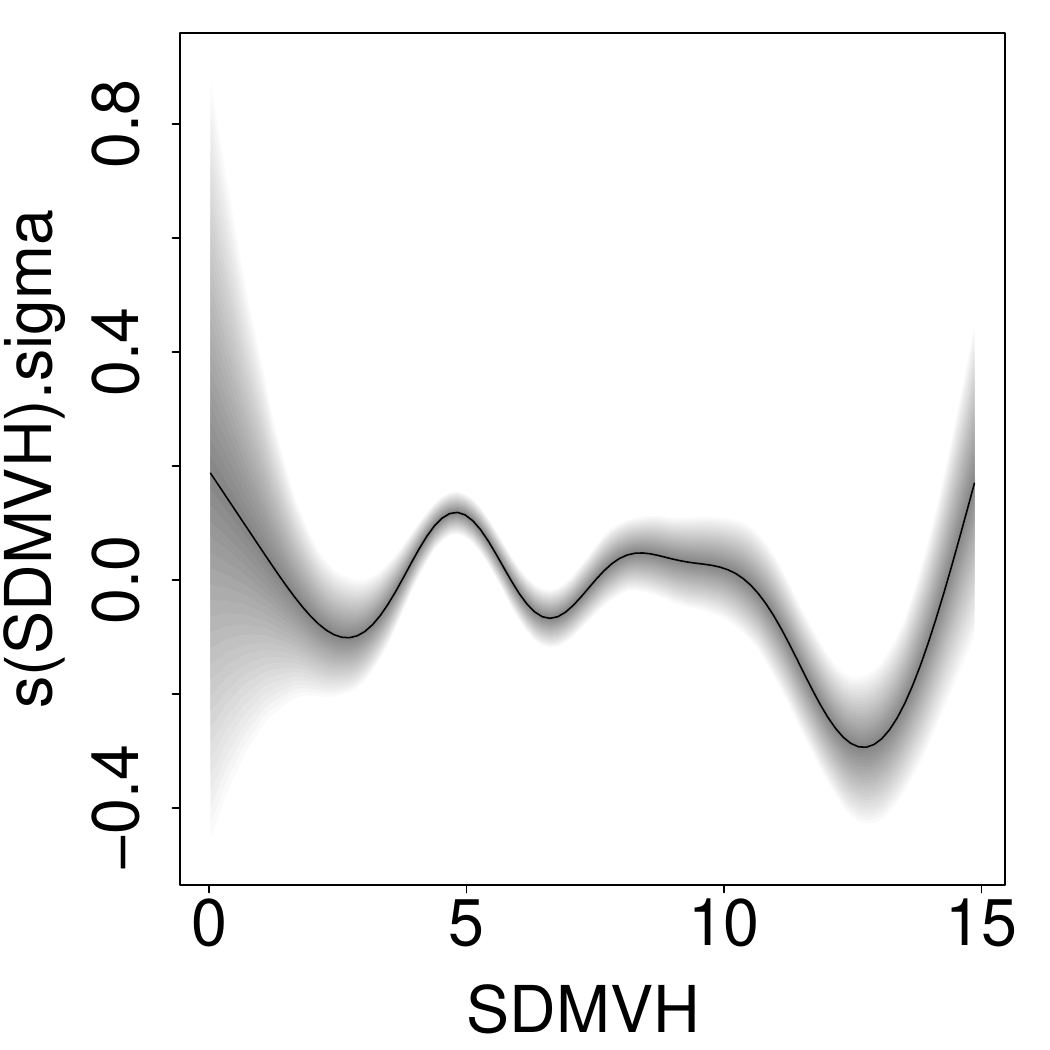}\\ 
\includegraphics[width=1\textwidth]{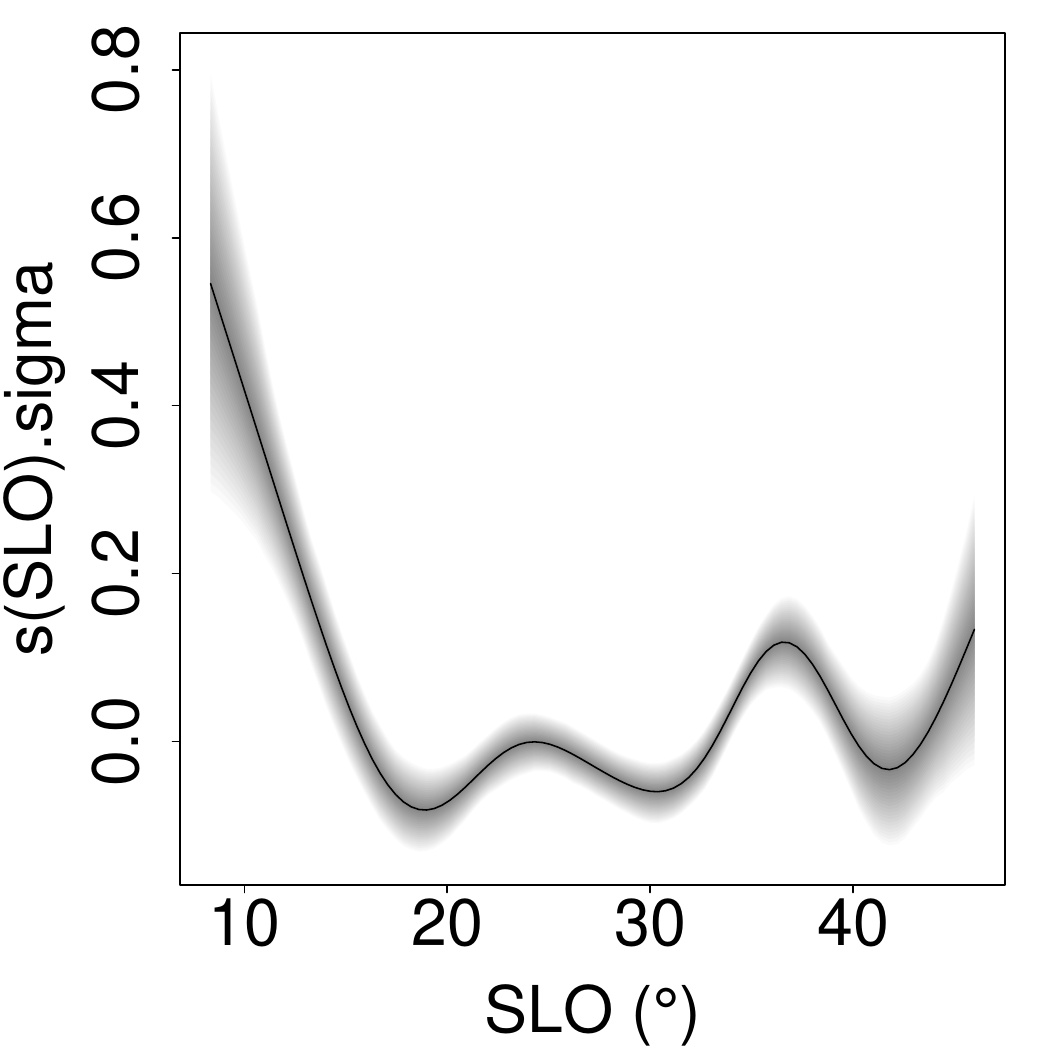}\\
\includegraphics[width=1\textwidth]{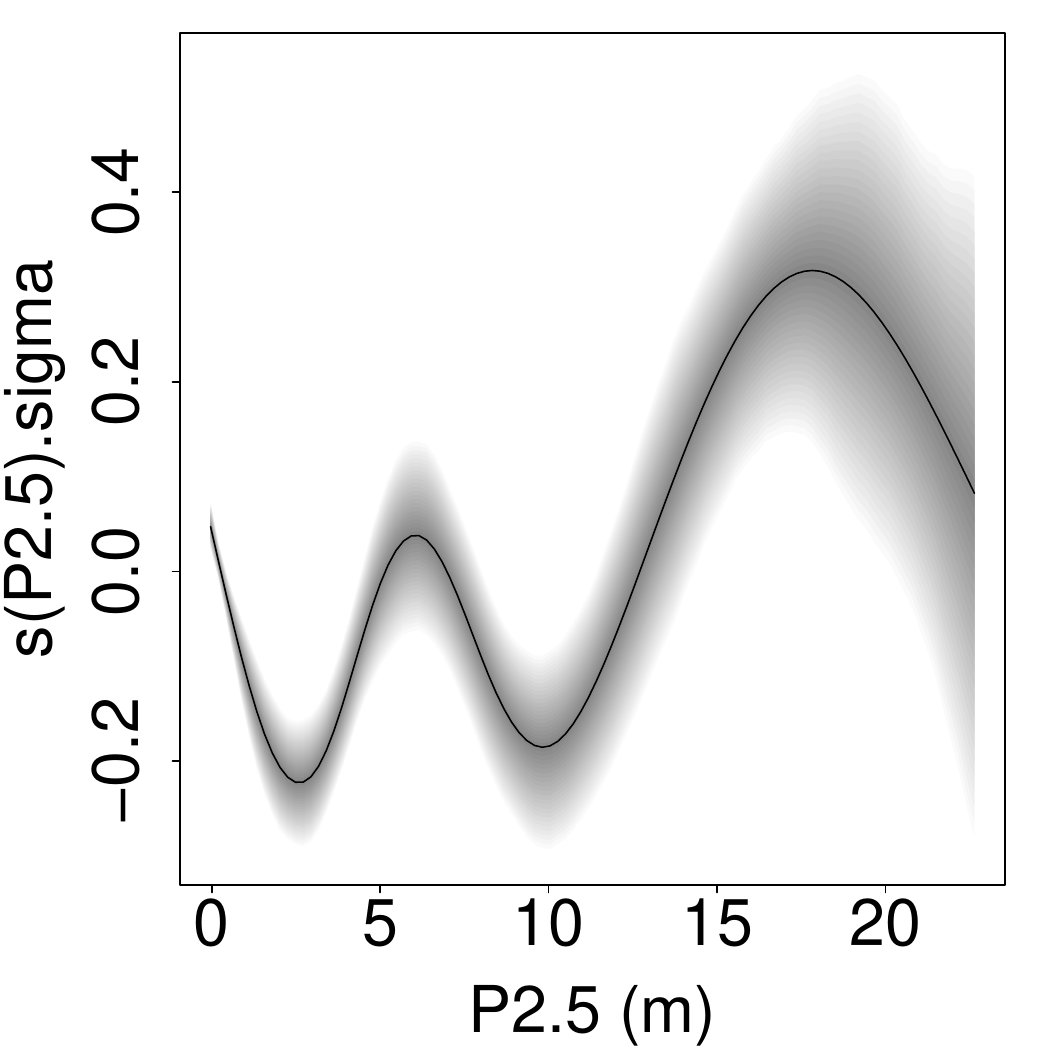}\\
\includegraphics[width=1\textwidth]{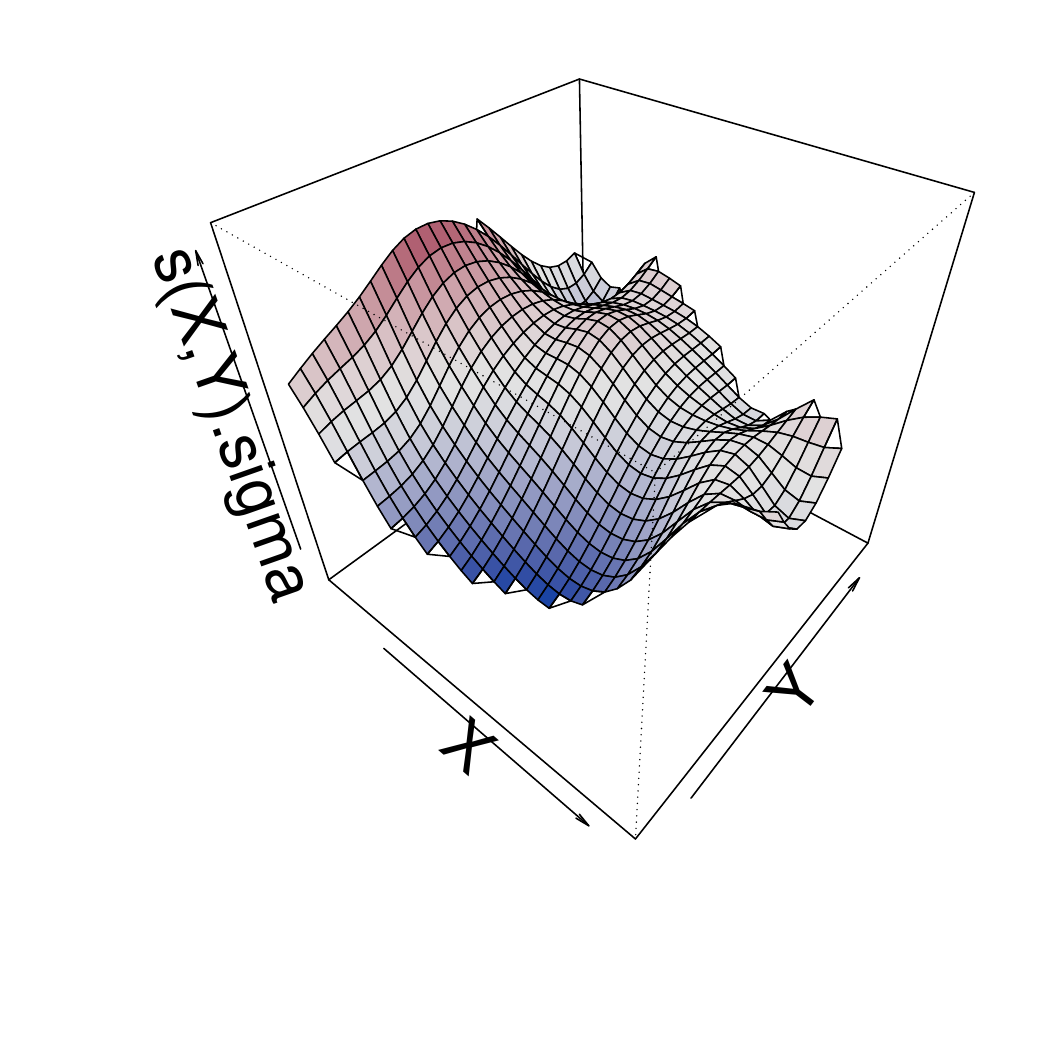}
\end{minipage}
\caption{Effect curves and 95\% credible intervals of the ``best'' model m\_14. Covariate effects on $\mu$ were indicated by s($\cdot$).mu, and effects on $\sigma$ by s($\cdot$).sigma.}
\label{fig:effect_curves}
\end{figure*}

The effect curves of model m\_14 (Fig.\,\ref{fig:effect_curves}) showed the mean vegetation height (MVH) and elevation above sea level (ESL) had positive effects on $\mu$ and $\sigma$ parameters. The slope of the terrain (SLO) as well as the 97.5-$th$ percentile of the pixellated vegetation height measures (P97.5) had almost strictly negative effects on $\mu$ and $\sigma$. The standard deviation of the vegetation heights (SDMVH) had a strictly positive effect on $\mu$. However, the effect of SDMVH on $\sigma$ behaved ambiguous for values below 10, had a negative effect between 10 and 13, and acted positively for values greater than 13. The cyclic effect of the topographic aspect (ASP) on $\mu$ had a local minimum at 150$^\circ$ (southeast) and a local maximum around 200$^\circ$ (south-southwest). The cyclic ASP effect on $\sigma$ had two local maxima at 75$^\circ$ (east-northeast) and 150$^\circ$ (southeast), and it had two local minima at 170$^\circ$ (south) and 280$^\circ$ (west). The effect of the 2.5-th percentile of the pixellated vegetation heights (P2.5) was indistinct for values less than 10\,m, but for greater values, it had a positive effect on $\mu$ and on $\sigma$.

\begin{figure*}
\centering
\includegraphics[width=0.5\textwidth]{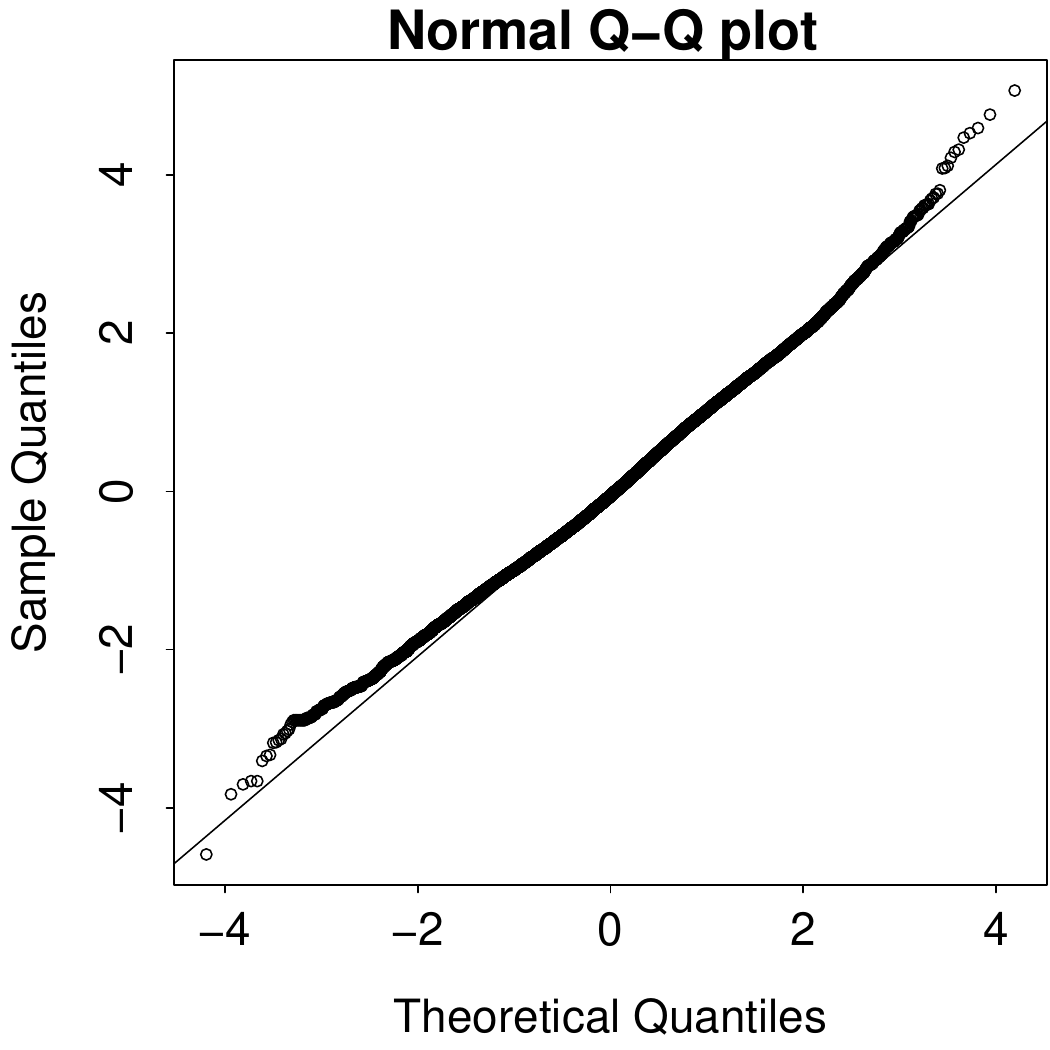}
\caption{Quantile-quantile plot of the residuals from model m\_14.}
\label{fig:qq_plot}
\end{figure*}

The quantile-quantile plot (qq-plot) in Fig.\,\ref{fig:qq_plot} shows quantile residuals lay close to the bisecting line between -2 and 3. This suggests the gamma distributional assumptions fit very well to the data and the distributional regression model m\_14 was adequately specified.

\begin{figure*}
\centering
\includegraphics[width=0.5\textwidth]{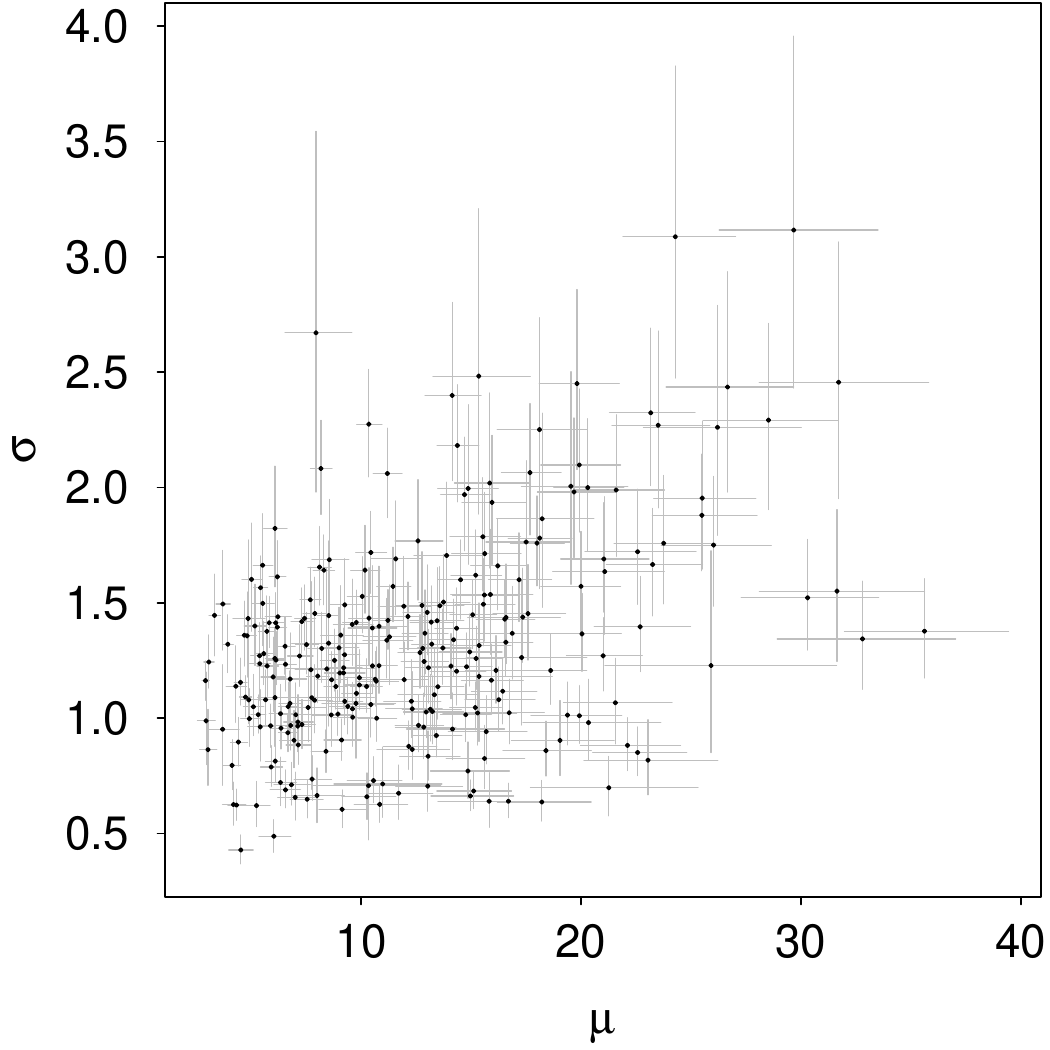}
\caption{Posterior estimates of $\mu$ and $\sigma$ for the model data from the 237 sample plots. The dots indicate the posterior mean, and the segments show 95\% credible intervals.}
\label{fig:mu_sigma_sampleplots_posterior}
\end{figure*}

For the model data of the 273 sample plots, the posterior mean estimates from the $M$ MCMC samples of the $\mu$ parameter ranged between 2.93 and 35.62, and the 273 posterior mean estimates of $\sigma$ parameter lay between 0.43 and 3.1 (Fig.\,\ref{fig:mu_sigma_sampleplots_posterior}). The correlation between the 273 posterior mean estimates for $\mu$ and $\sigma$ was 0.42. The average relative mean squared error (MSE\%) of the $\mu$ estimates was 5.0\%, and the $\sigma$ estimates had a MSE\% of 7.2\%.

Empirical histograms and posterior distributional predictions of the DBH distributions on the 273 model data sample plots are presented in Figs.\,S1--S8 of the supplemental material. These figures show the distributional predictions fit very well to the empirical histograms across all sample plots.

\begin{figure*}
\begin{minipage}[c]{0.49\textwidth}\centering
\includegraphics[width=0.9\textwidth]{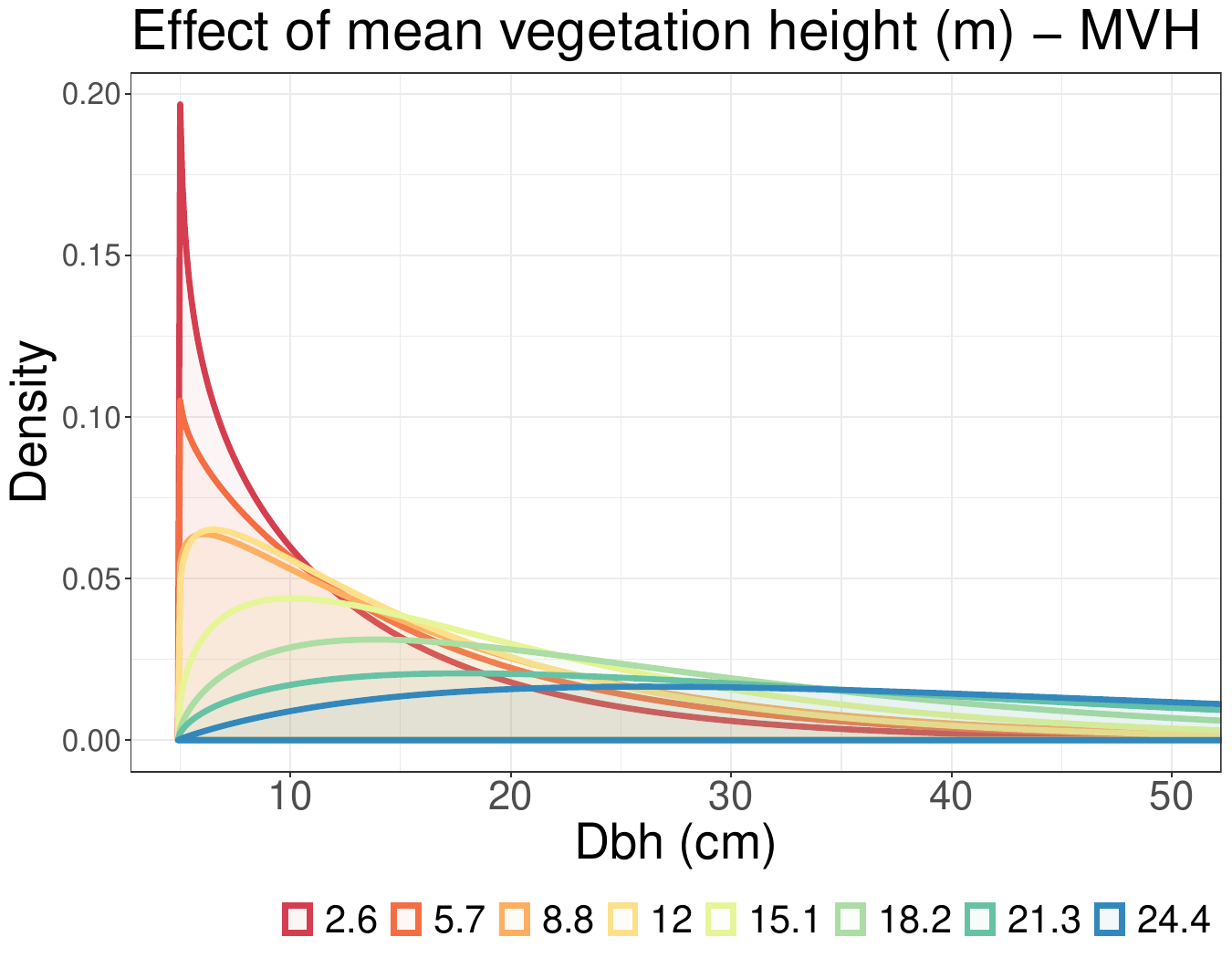}\vspace*{10pt}\\ 
\includegraphics[width=0.9\textwidth]{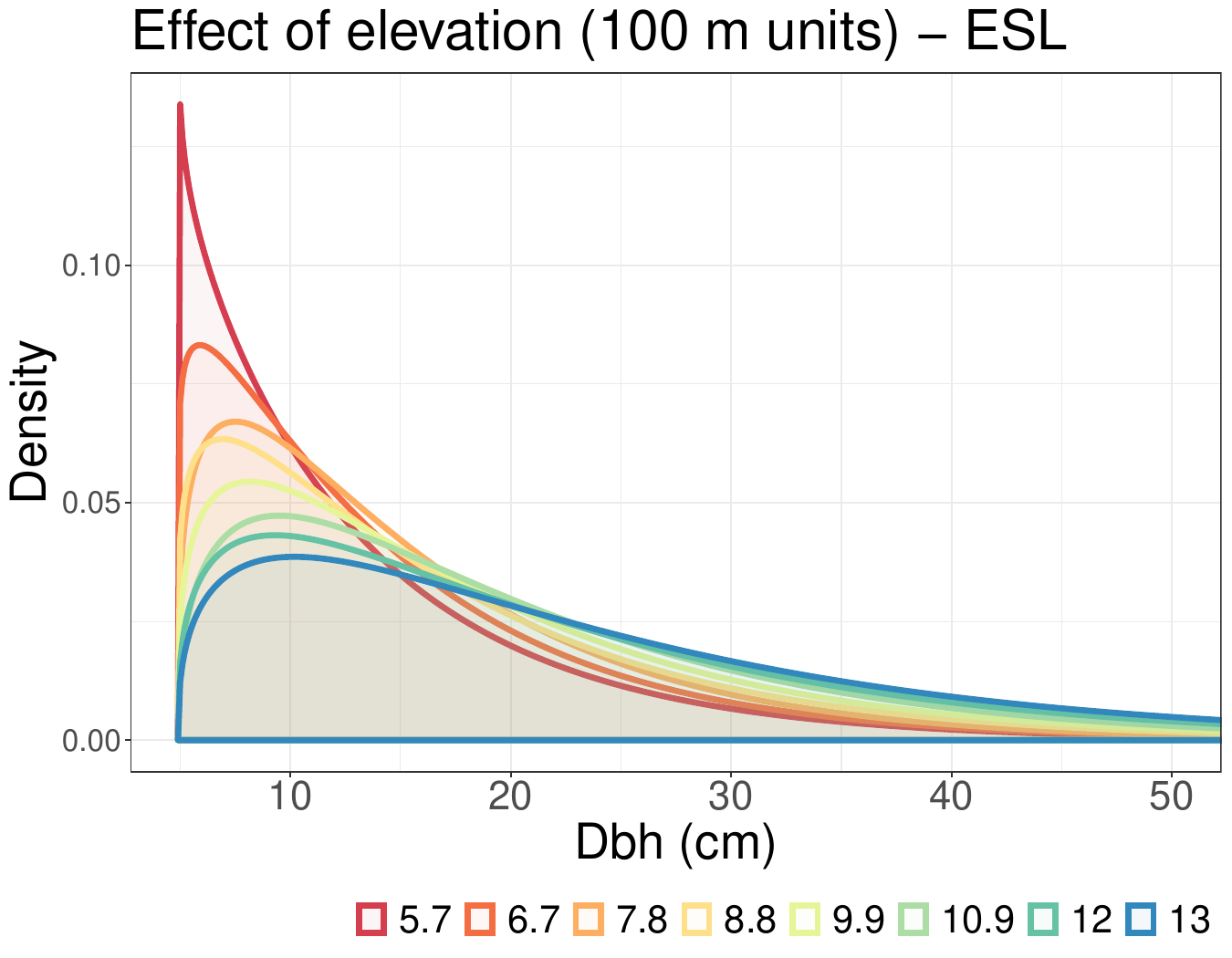}\vspace*{10pt}\\
\includegraphics[width=0.9\textwidth]{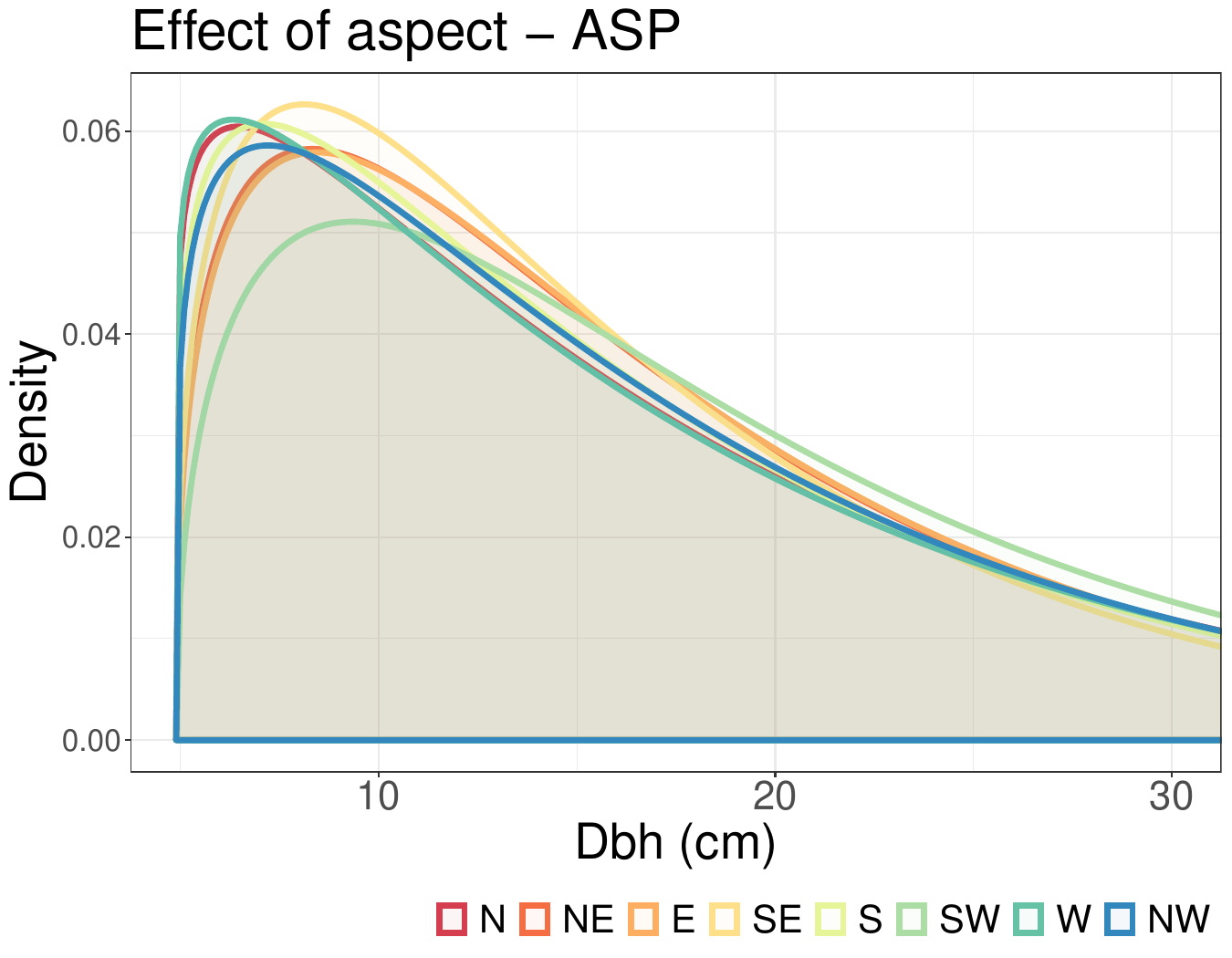}
\end{minipage}
\begin{minipage}[c]{0.49\textwidth}\centering
\includegraphics[width=0.9\textwidth]{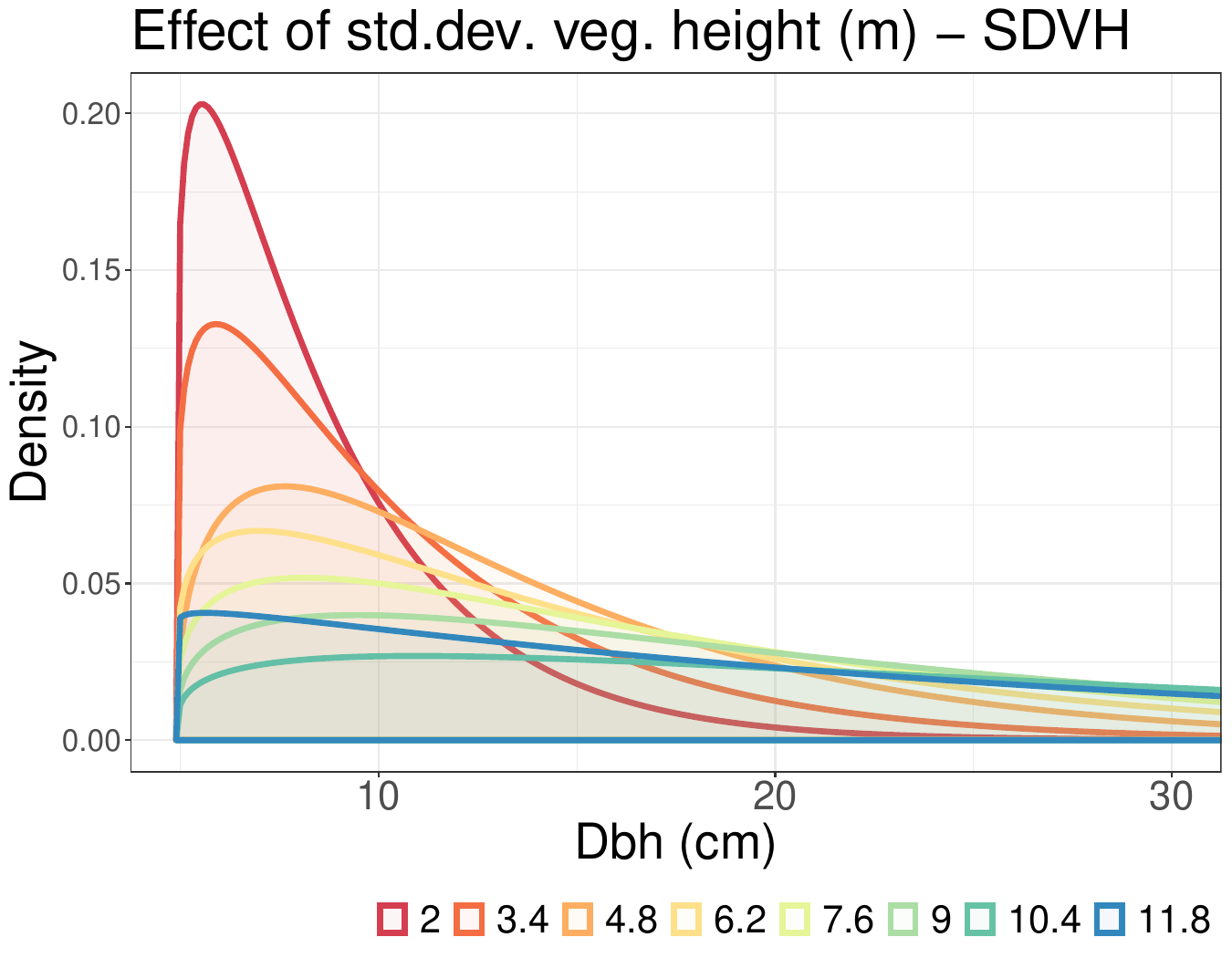}\vspace*{10pt}\\ 
\includegraphics[width=0.9\textwidth]{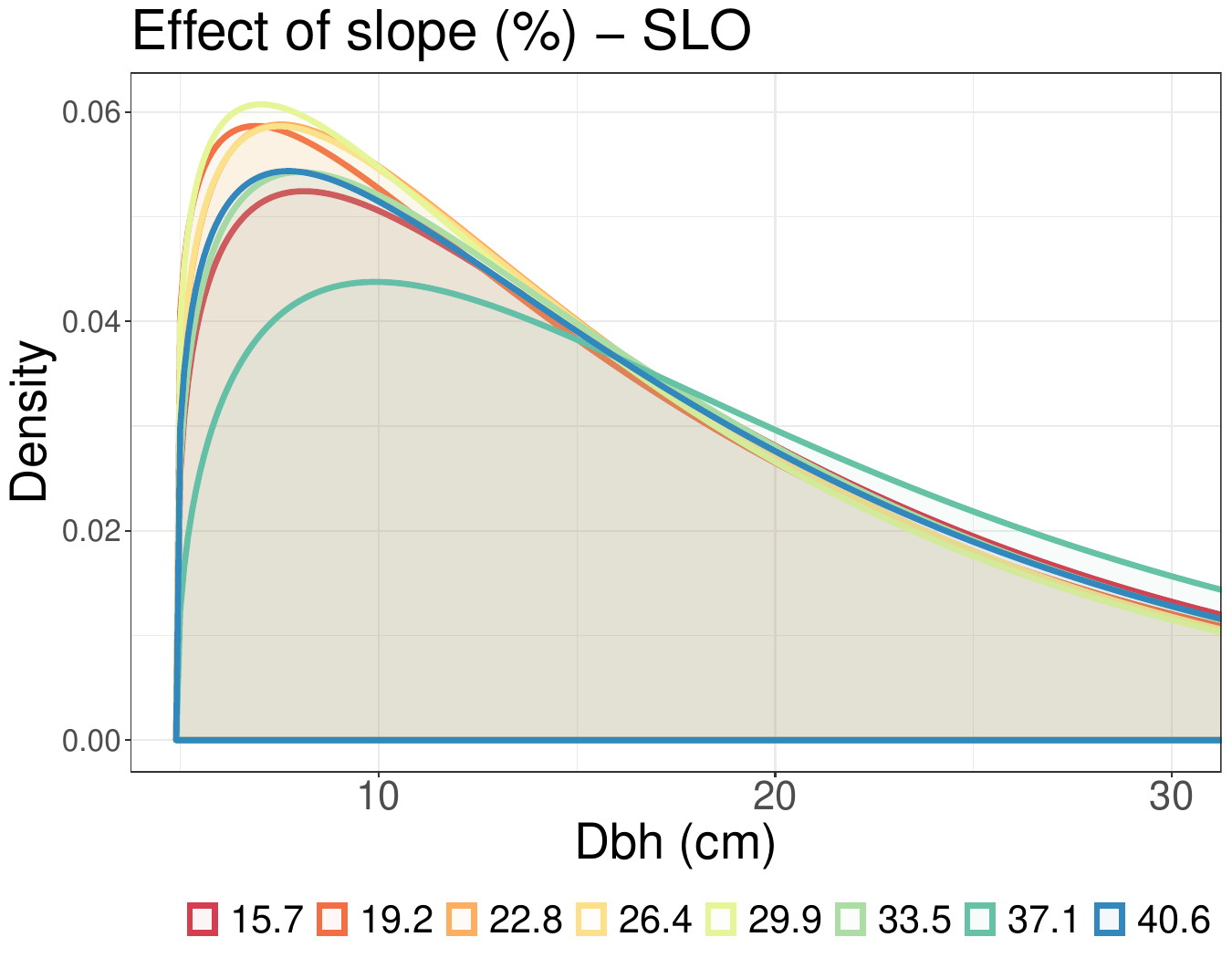}\vspace*{10pt}\\
\includegraphics[width=0.9\textwidth]{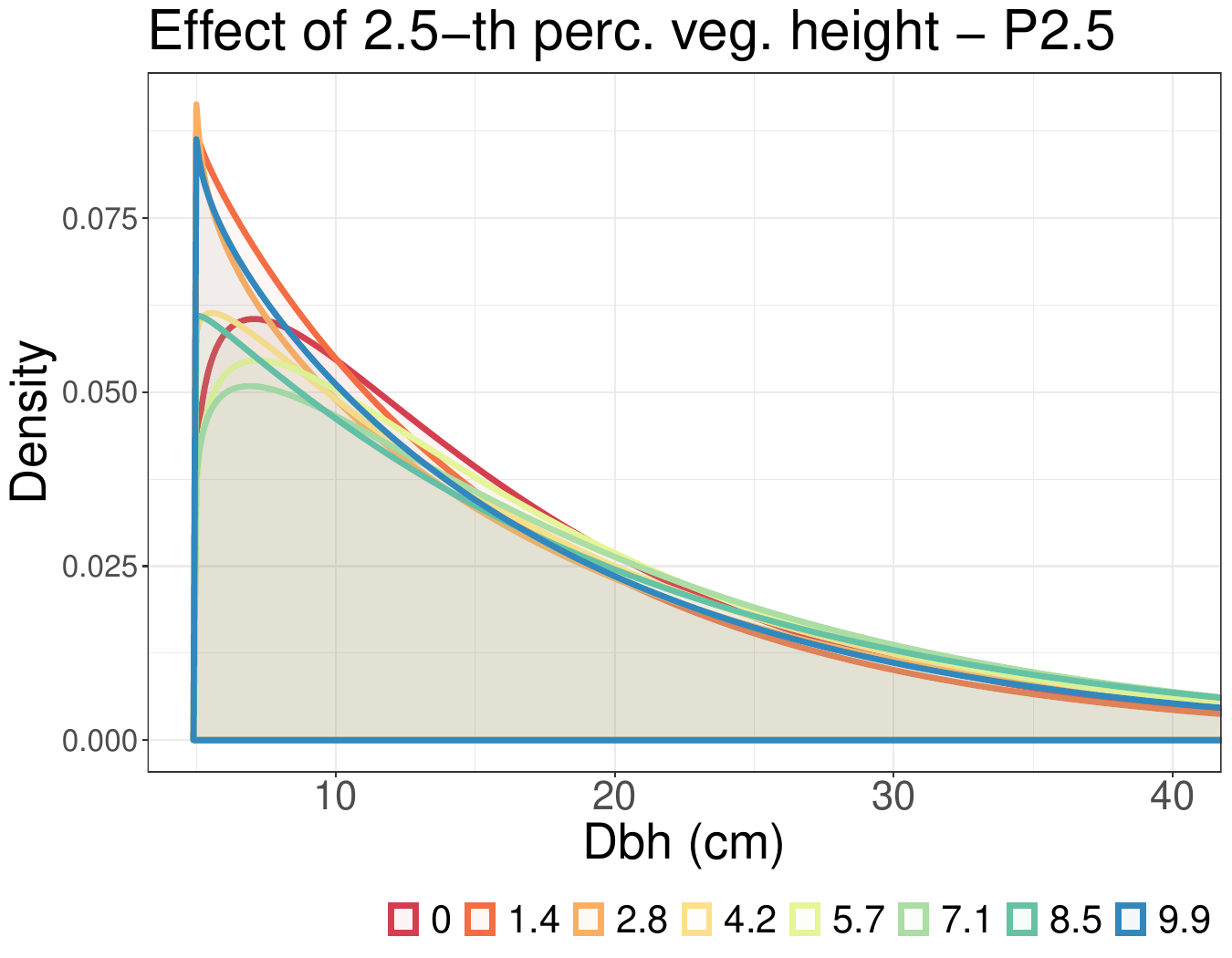}
\end{minipage}\vspace*{10pt}\\
\centering
\includegraphics[width=0.441\textwidth]{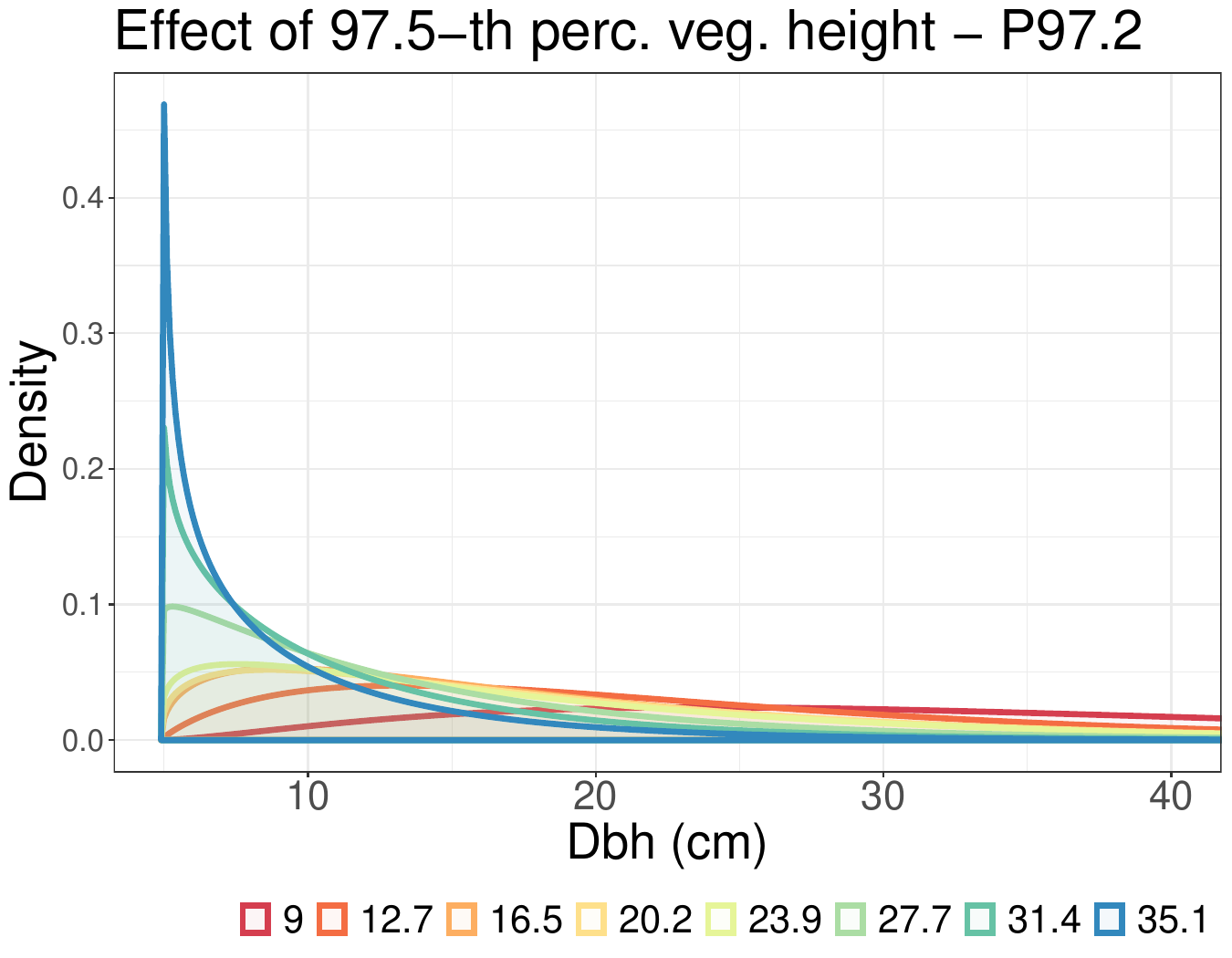}
\caption{Visualization of the covariate effects on the DBH distribution.}
\label{fig:dist_vis}
\end{figure*}

To assess what influence the covariates simultaneously had on both distribution parameters $\mu$ and $\sigma$, the gamma density was evaluated under \emph{ceteris paribus} conditions for grid values within the range of a single covariate, while the other covariates were kept fixed at their respective median values (Fig.\,\ref{fig:dist_vis}). As MVH acted positively on the expectation as well as on the variance of the gamma distribution, an increasing MVH flattened the density and shifted the mass towards higher DBH values. Similar effects occurred for increasing SDVH and ESL values. A completely opposite effect became obvious for an increasing P97.5. More complex and nonlinear effects on the DBH distribution were observed for SLO, ASP, and P2.5.

\subsection{Distributional prediction}\label{sec:distributional_prediction}

\begin{figure*}\centering
\includegraphics[width=1.0\textwidth]{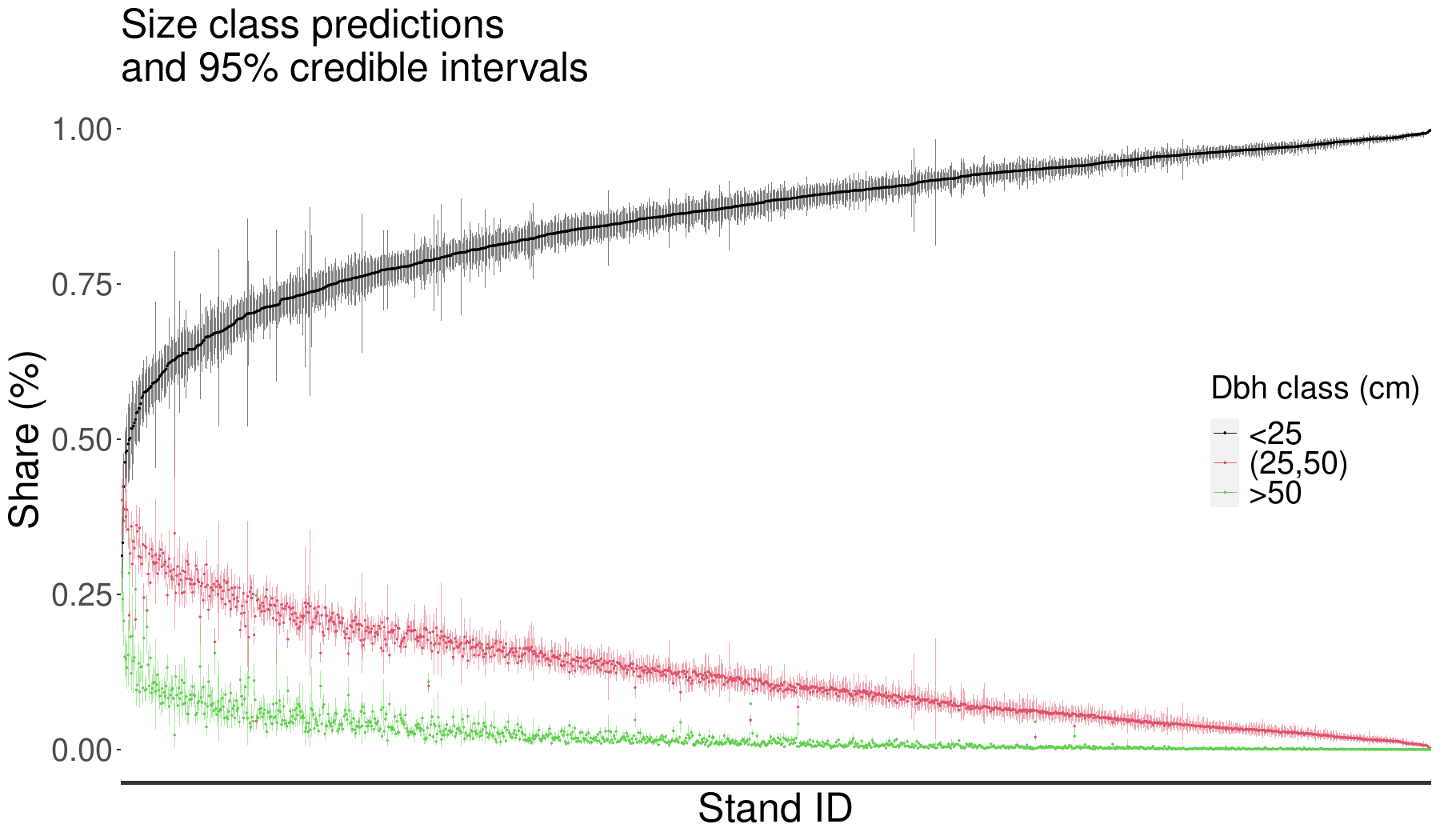}
\caption{Predictions of DBH class allocations across the forest stands for the two classification variants.}
\label{fig:size_class_prediction}
\end{figure*}

As noted previously, the Ebensee forest district domain was partitioned into 35.5\,m$\times$35.5\,m pixels. For each of these prediction pixels, covariate data were derived from the DVHM and the DSM in terms of the MVH, SDMVH, ESL, SLO, ASP, P2.5, P97.5, and the pixel centroid coordinates. The MCMC samples from the posterior parameter distributions for the covariate effects of model m\_14 were then applied with these covariates to achieve posterior predictive distributions of $\mu$ and $\sigma$ for each prediction pixel. Consequently, the gamma distribution was evaluated using these parameter estimates to produce predictions of stem count proportions that fall into the DBH classes such as specified in Section\, \ref{sec:prediction}. Finally, an area-weighting scheme was applied to achieve aggregated predictions of these stem count proportions throughout all prediction pixels per forest stand. Such as demonstrated in Fig.\,\ref{fig:size_class_prediction}, the 95\% credible intervals were relatively tight and the size class predictions became highly precise across all forest stands. 

\begin{figure*}
\begin{minipage}[c]{0.49\textwidth}\centering
\includegraphics[width=1.0\textwidth]{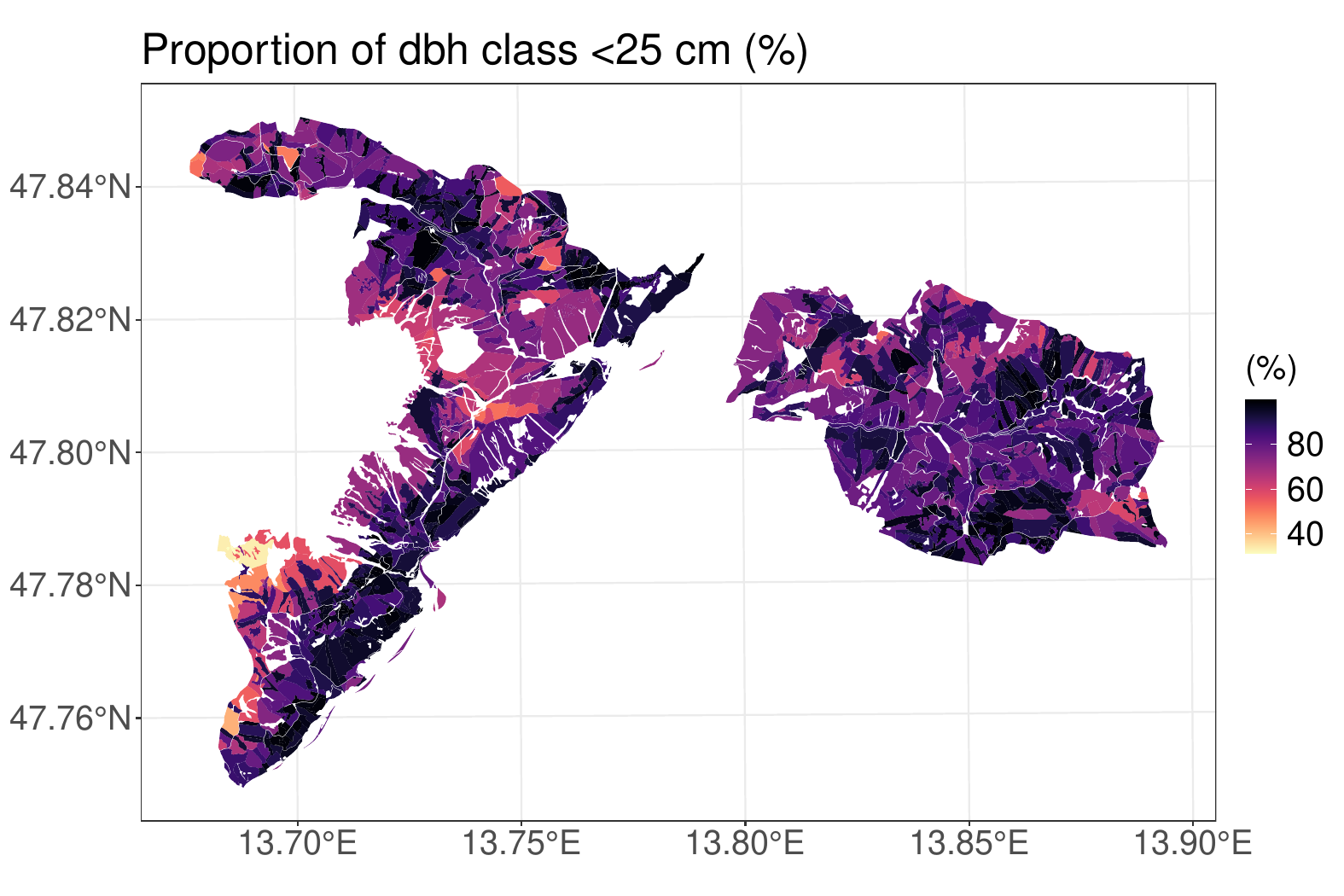}\vspace*{10pt}\\ 
\includegraphics[width=1.0\textwidth]{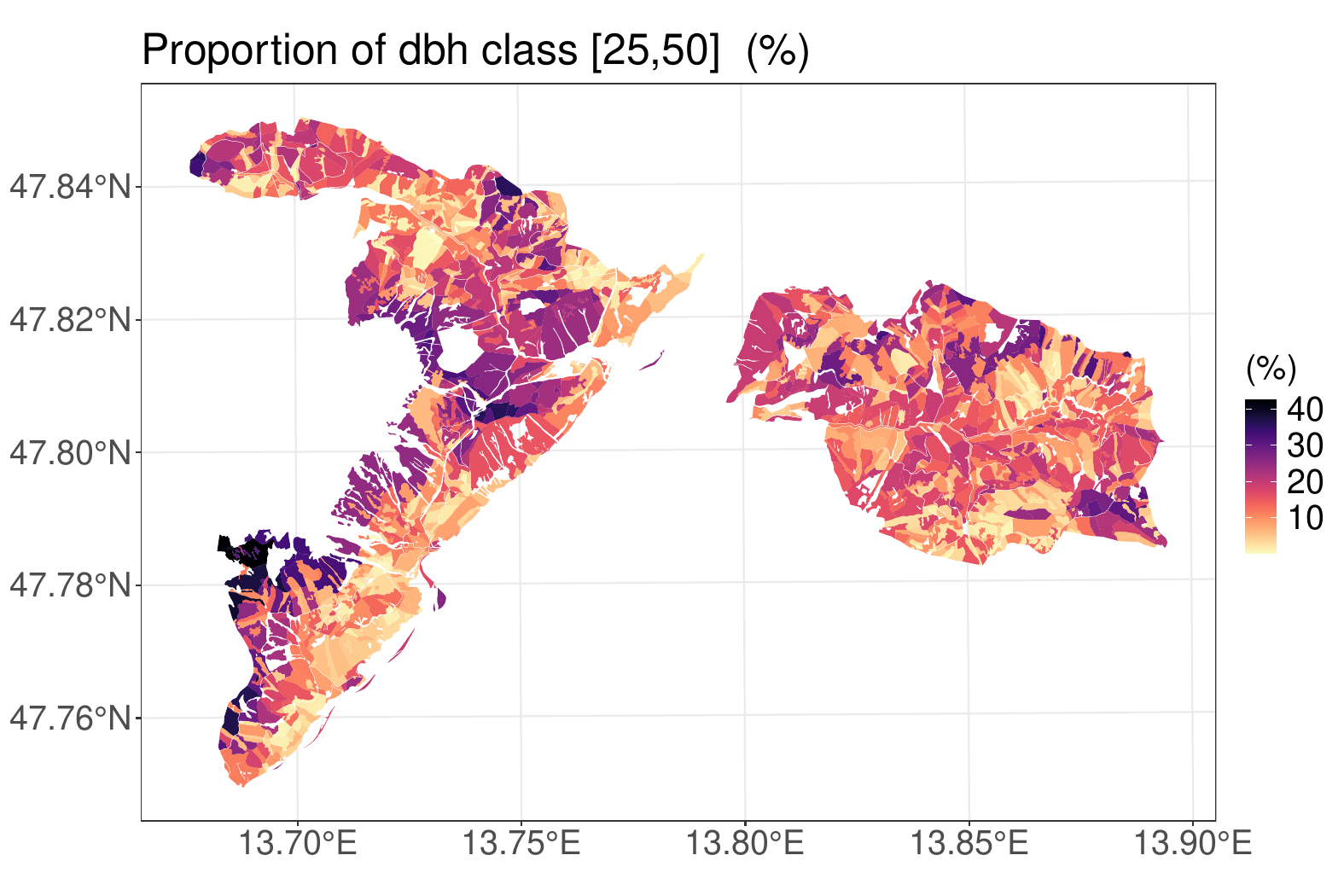}\vspace*{10pt}\\
\includegraphics[width=1.0\textwidth]{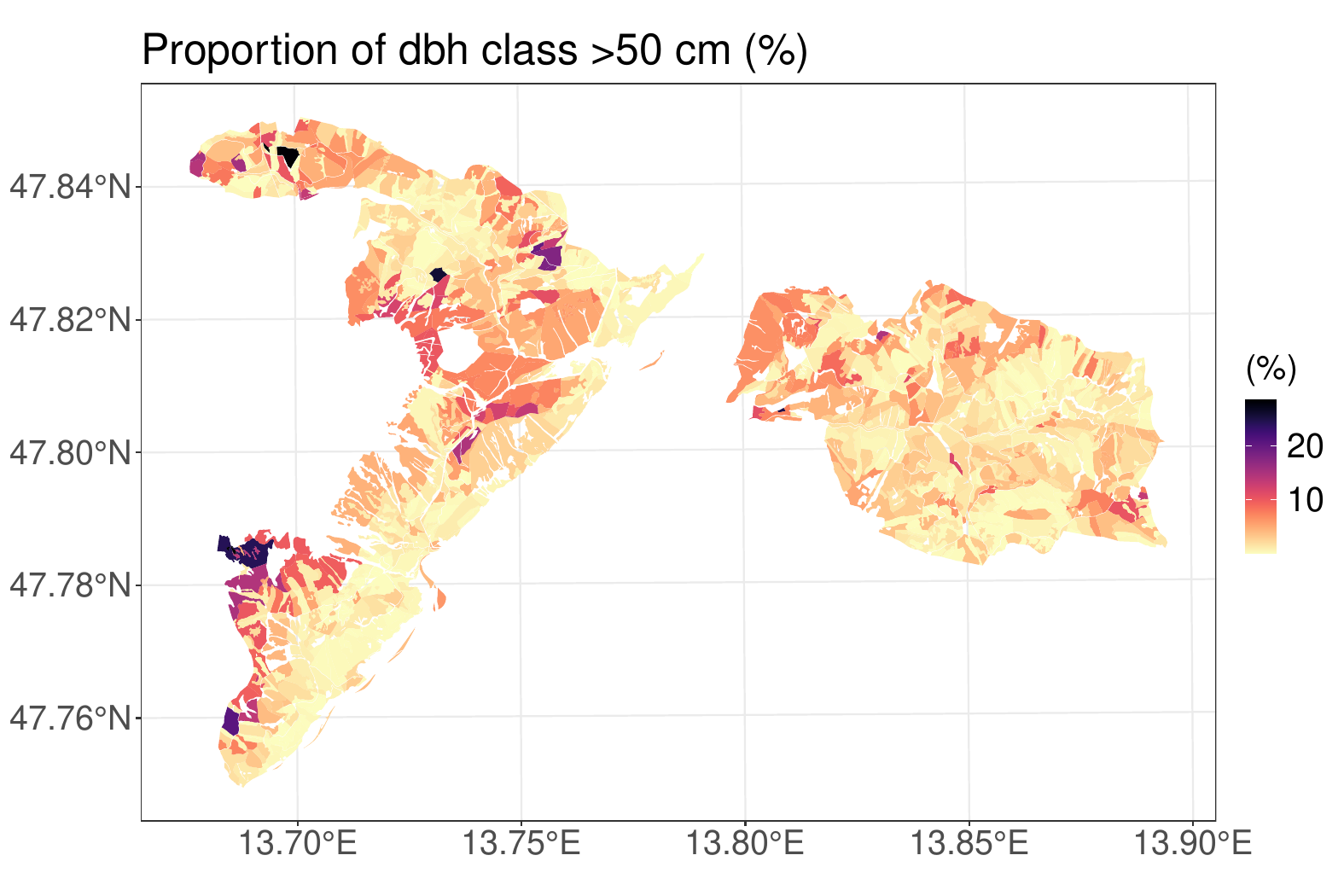}
\end{minipage}
\begin{minipage}[c]{0.49\textwidth}\centering
\includegraphics[width=1.0\textwidth]{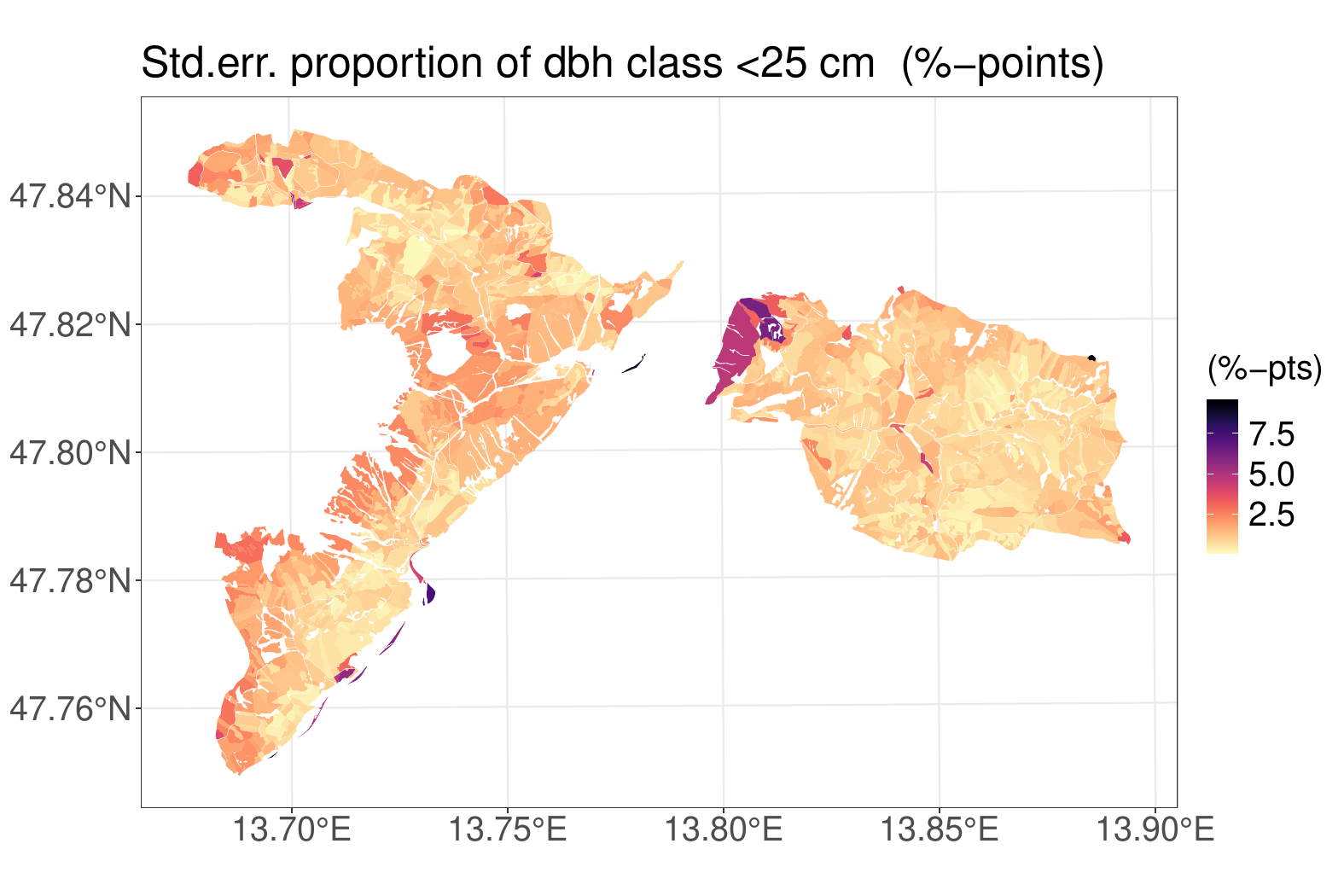}\vspace*{10pt}\\ 
\includegraphics[width=1.0\textwidth]{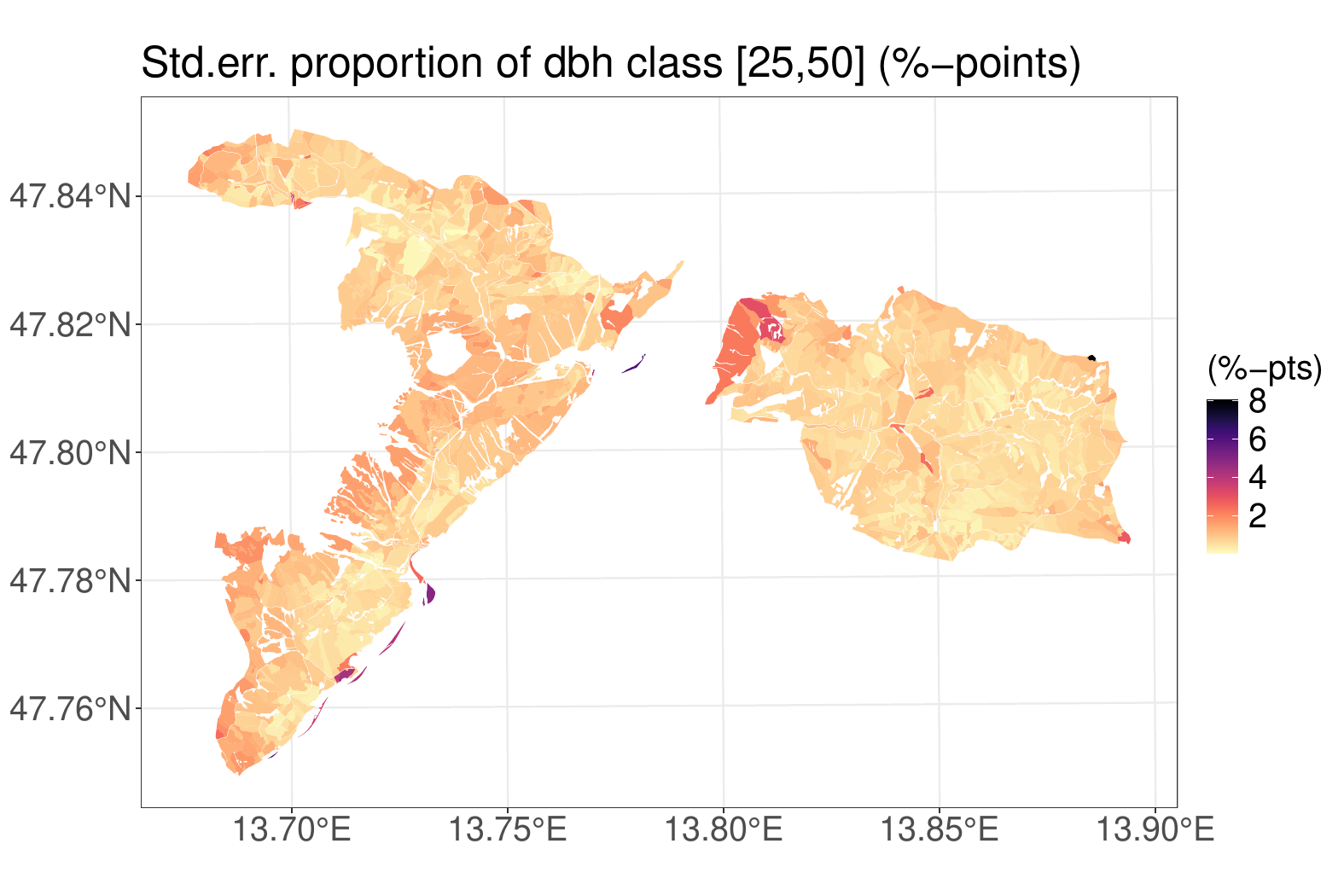}\vspace*{10pt}\\
\includegraphics[width=1.0\textwidth]{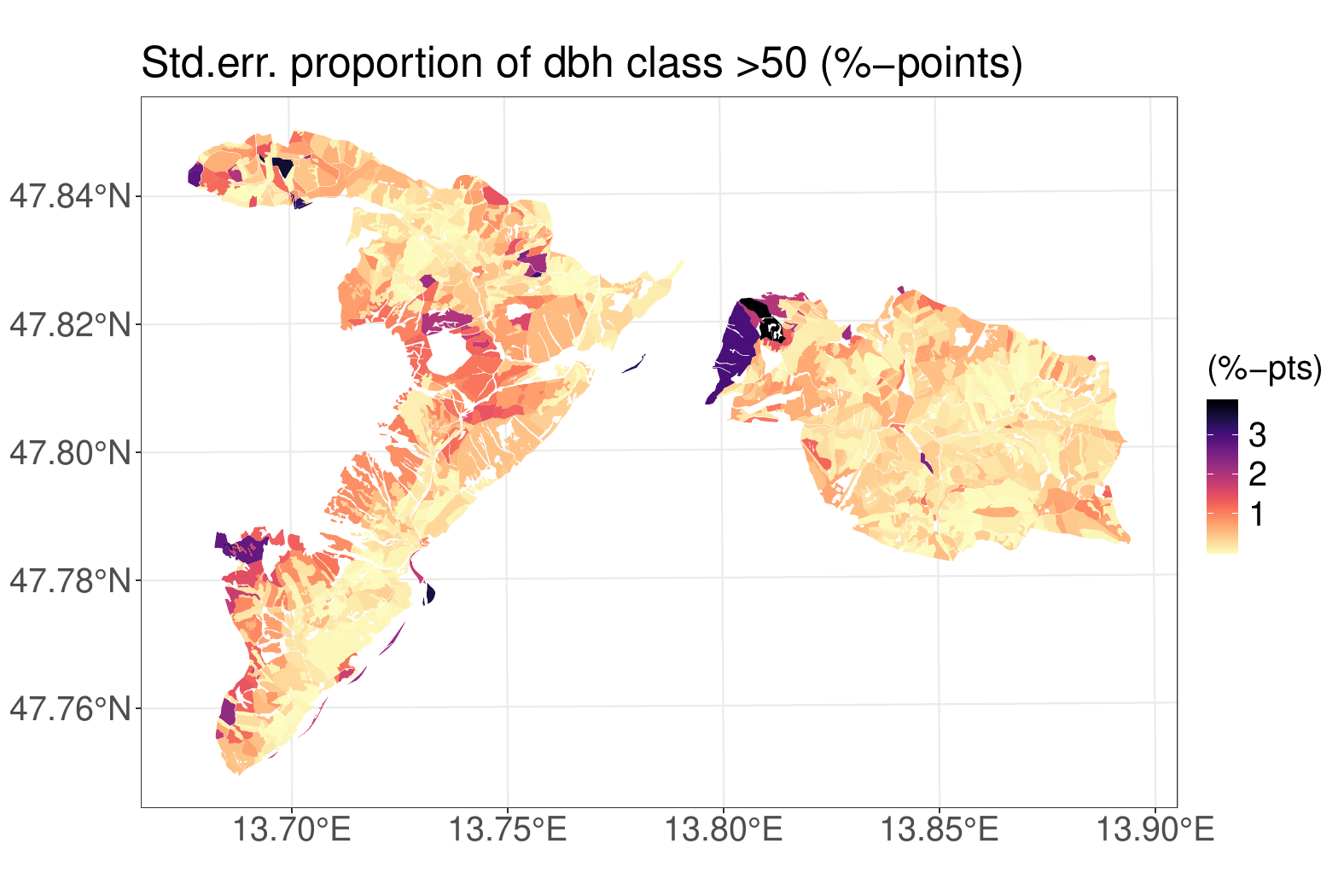}
\end{minipage}\vspace*{10pt}\\
\caption{Spatial size class predictions (\%) and mean squared error (\%-points) for the forest stands in the forest distric Ebensee using the classification: (1) small (DBH$<$25\,cm), (2) intermediate (25\,cm$\leq$DBH$\leq$50\,cm), and (3) large (DBH$>$50\,cm).}
\label{fig:spatial_sizeclass_pred_25_50}
\end{figure*}

Maps of these size class predictions (Fig.\,\ref{fig:spatial_sizeclass_pred_25_50}) revealed that smaller tree sizes were especially lacking in the central area of the Feuerkogel region located in the western part of the forest district Ebensee, while these stands also possessed a relatively high proportion of larger trees. Such as reported by the sample plot field crew, these sites were actually in a mature state, and establishment of natural regeneration was hindered so far by the dense shelter of larger mature trees. Appropriate silvicultural management activities are therefore needed to restore the protection function in this area. The size class predictions were relatively precise for both classification schemes and across all forest stands. The average MSE was only 0.85 percentage points, and the maximum was 9.6. The MSE was less than 1.6 percentage points for 90\% of the size class predictions, and 95\% of the predictions had a MSE less than 2 percentage points.


\begin{figure*}\centering
\includegraphics[width=0.5\textwidth]{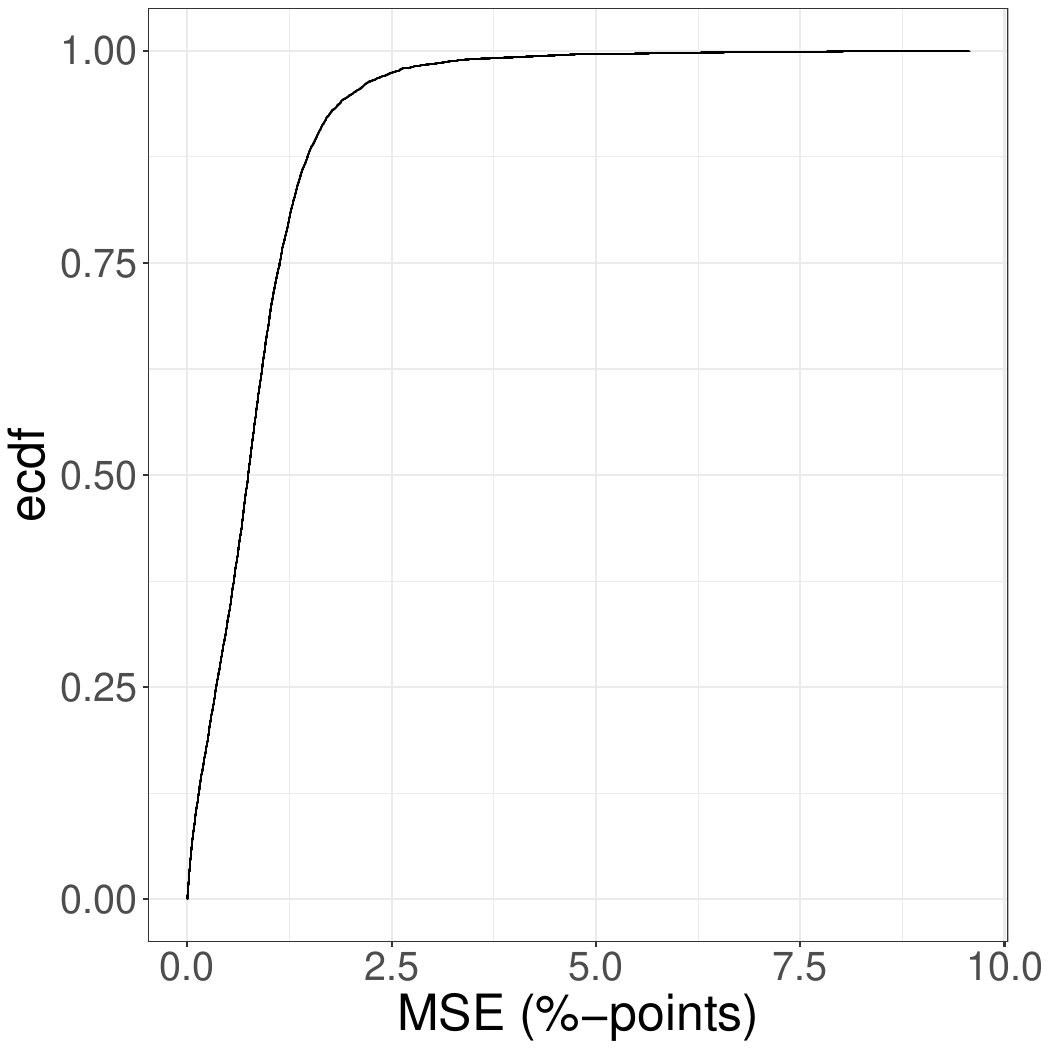}
\caption{Empirical cumulative distribution function (ecdf) of the size class prediction mean squared errors (MSE).}
\label{fig:ecdf_mse}
\end{figure*}

\section{Discussion}\label{sec:discussion}

Posterior standard deviations of the DBH class predictions were considerably low due to the spatially dense network of PLS sample points. Some areas of higher uncertainty are in the top corner of the Eastern half, which were characterized by irregular forests on steep slopes, and due to rock faces no PLS sample points could be captured. Also the South-Western ridge in the Western half has higher errors, caused by missing points due to inaccessibility and generally ragged terrain in high altitudes with less forest cover. 

Spatially coherent diameter distribution predictions and subsequently derived probabilistic maps of meaningful size classes provided useful tools to support forest management decisions. In the Ebensee area we found an overall high proportion of small diameter trees, which are essential to sustain the protective function, especially in that steep terrain. Due to a dense forest road network in the forest district it is relatively easy to maintain forest regeneration. Nevertheless some regions might need additional active management, especially in the central Western half and in the Eastern half the central-top as well as the South-Eastern corner. Predictions of the diameter distribution alone are still insufficient to fully assess the forest's protective function, and a special interest would be in assessing the structural change over time.  

The proposed modeling framework is flexible and able to represent all the structural differences among the sample plots. However, forest stands could theoretically possess a layered structure with an understorey of younger and thinner trees growing under a shelter of older and thicker trees. These circumstances often result in a multimodal DBH distribution. Practically, such structure could be modeled through a mixture of two or more different density functions. However, the methodology so far provided by the \verb|BAMLSS| package is restricted to a single density and does not allow construction of composed mixture models. This limitation was practically irrelevant for our study, because none of the sample plots showed clear signs of a strict multimodal DBH distribution; compare Figs.\,S1--S8 in the supplemental material.

In addition, any problems that could have been caused by \verb|BAMLSS|'s limitation to unimodal distributions were ameliorated by our approach to achieve the predictions at forest stand level. Hereby, the forest stands were partitioned into smaller prediction pixels. For each of the prediction pixels, the gamma distribution parameters were predicted, and the density was evaluated. The forest stand level prediction of the DBH distributions was finally achieved via an area-weighted aggregate of the pixel-level densities. In so doing, the final prediction of the DBH distribution at stand level was no longer restricted to a unimodal parametric shape.

Forest inventory field work was conducted using a PLS to create ``digital twins'' of the vegetation and the terrain on a 20\,m radius plot. The field work was very efficient with approximately 12 minutes labor time per plot, including the set-up of the equipment and the scanning process. By using fully automated routines, 133 trees were measured on average per sample plot, with LiDAR-derived information being not only DBH, but also other parameters such as height or crown base. This is in contrast to the traditional forest inventory practice, in which  measurements are conducted with optical and mechanical instruments. Because these instruments consume high labor costs, traditional forest inventory uses much smaller plot sizes than our 20\,m radius plots, so that often not more than 10 trees were measured per plot. With such smaller sample sizes of the traditional forest inventories, the distributional regression modeling would have been hardly possible, and the novel PLS-supported forest inventory can be regarded as key to successful DBH distribution modeling and prediction.

In this study, an approach was presented to model and predict stem diameter distributions in terms of a parametric probability density function. To produce a quantitative prediction of the absolute stem count per DBH class, a further estimate of the total stem count per area unit is needed. A possible approach to achieve such estimate would be to couple the proposed spatial distribution regression model with an extra spatial spatial regression model that considers the tree count per hectare as response. Appropriate methodology for the spatial regression modeling of the growing stock timber volume per area unit is presented in \cite{Nothdurft2021} and could be adopted to model number of trees per hectare. To date, high density ALS data is available for the complete Ebensee forest district domain and will probably be maintained in future, as the area is designated as research zone and has been of particular interest of the Austrian Federal Forest Service. In future work, we will therefore also test an individual tree segmentation from the ALS canopy height model using methodology implemented in the \verb|R|-package \verb|lidR| by \cite{Roussel2020, Roussel2023} that provides a comparative approach to the spatial regression model of tree counts.

\section{Conclusion}\label{sec:conclusion}

This study presented a method to estimate stem diameter distributions by linking PLS and ALS data in the protection forest landscape Ebensee. The Bayesian distributional regression framework was based on Gamma distributions as implemented in the \verb|BAMLSS| \verb|R|-package . The Gamma distribution's shape and scale parameters were modeled using linear predictors dependent on covariates from the PLS and ALS data.  \verb|BAMLSS| offered the modeling of nonlinear covariate effects by using penalized regression spline smoothers, which proved more favorable than linear parametric slope coefficients. Including spatially structured effects on both gamma parameters significantly enhanced the model performance. Thereby, the modeling of a spatial Gaussian process outperformed a bivariate tensor product smooth across the sample plot location coordinates.

A spatial wall-to-wall prediction of the gamma distribution was achieved by partitioning the entire domain into prediction pixels having an area equal to the sample plot. The DBH distributions were predicted at forest stand level via area-weighted aggregates of the evaluated posterior predictive densities.

The proposed model framework can be easily adopted to other tasks when information is required on forest structural diversity across broader forest landscapes. The latter aspect might be of special interest to forestry enterprises charged with protection forest management. In such settings, estimating DBH distributions and other forest structure measures can inform management decisions focused on sustaining protective forest characteristics.

\section{Acknowledgments}\label{sec:Acknowledgments}

This study was supported by the project Invent-PLS and was financed by the Austrian Federal Ministry of Finance via the the Austrian Research Promotion Agency (FFG) under project number FO999899975 and eCall number 47343364. S. Witzmann's work was completely financed by Invent-PLS. Finley's work was supported by Michigan State University AgBioResearch and NASA CMS grants Hayes (CMS 2020) and Cook (CMS 2018).

\bibliographystyle{apalike}
\bibliography{Distreg}

\newpage
\beginsupplement
\section*{Supplemental material}

\begin{figure*}
\centering
\includegraphics[height=1\textheight, page=1]{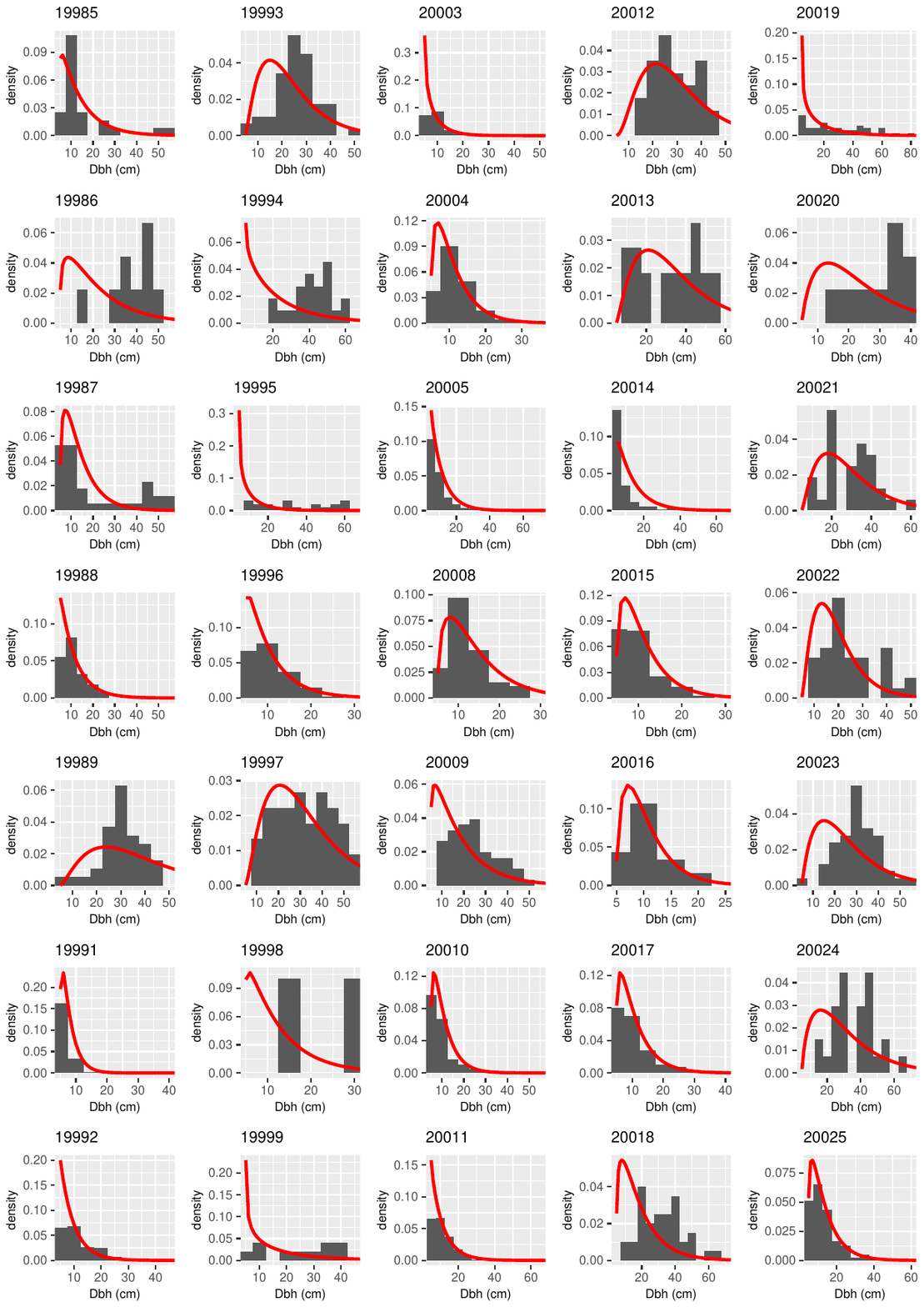}
\caption{Histograms and posterior predictive distributions of the DBH at the sample plots.}
\label{fig:dbh_pred_sampleplots_1}
\end{figure*}

\begin{figure*}
\centering
\includegraphics[height=1\textheight, page=2]{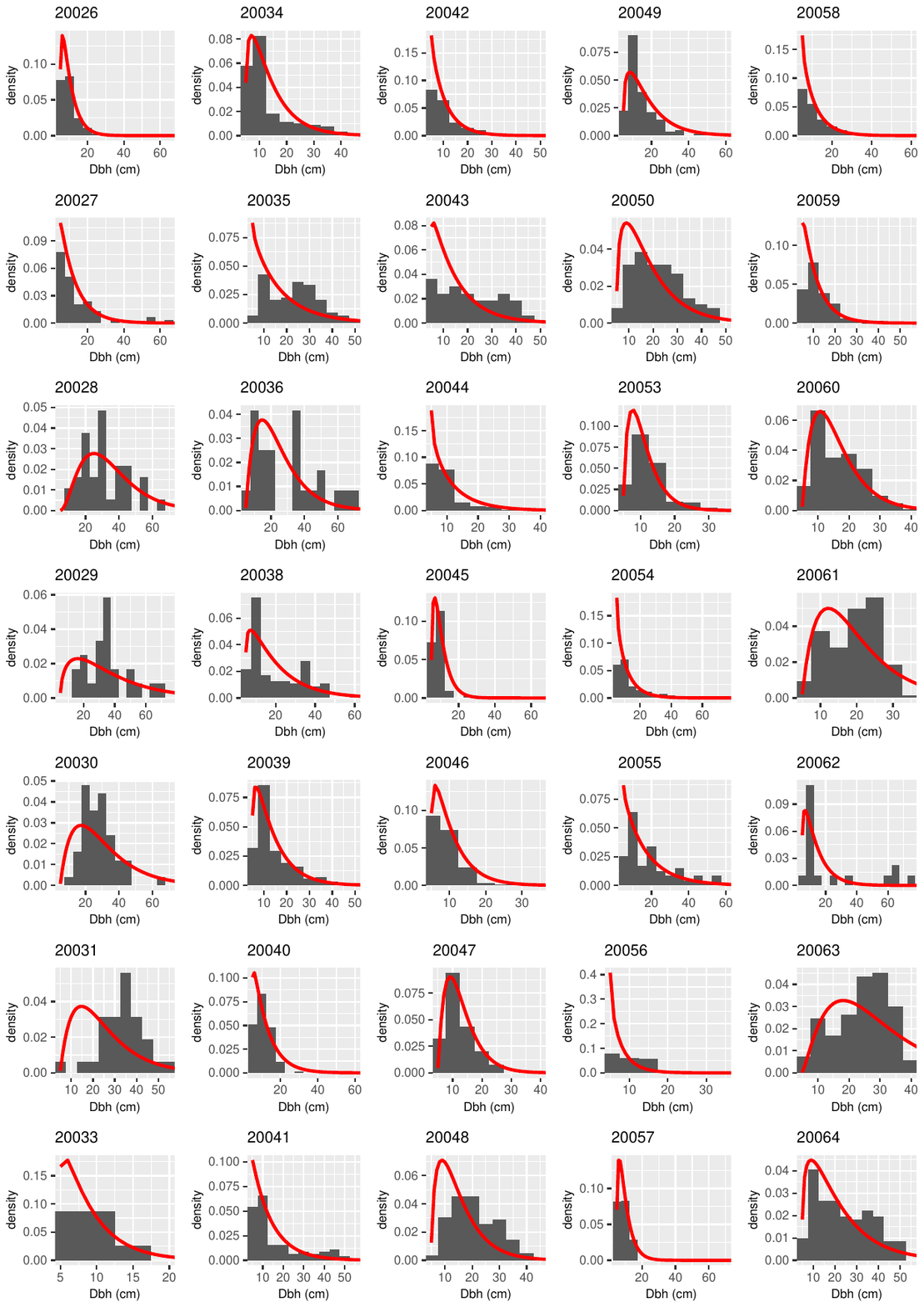}
\caption{Histograms and posterior predictive distributions of the DBH at the sample plots.}
\label{fig:dbh_pred_sampleplots_2}
\end{figure*}

\begin{figure*}
\centering
\includegraphics[height=1\textheight, page=3]{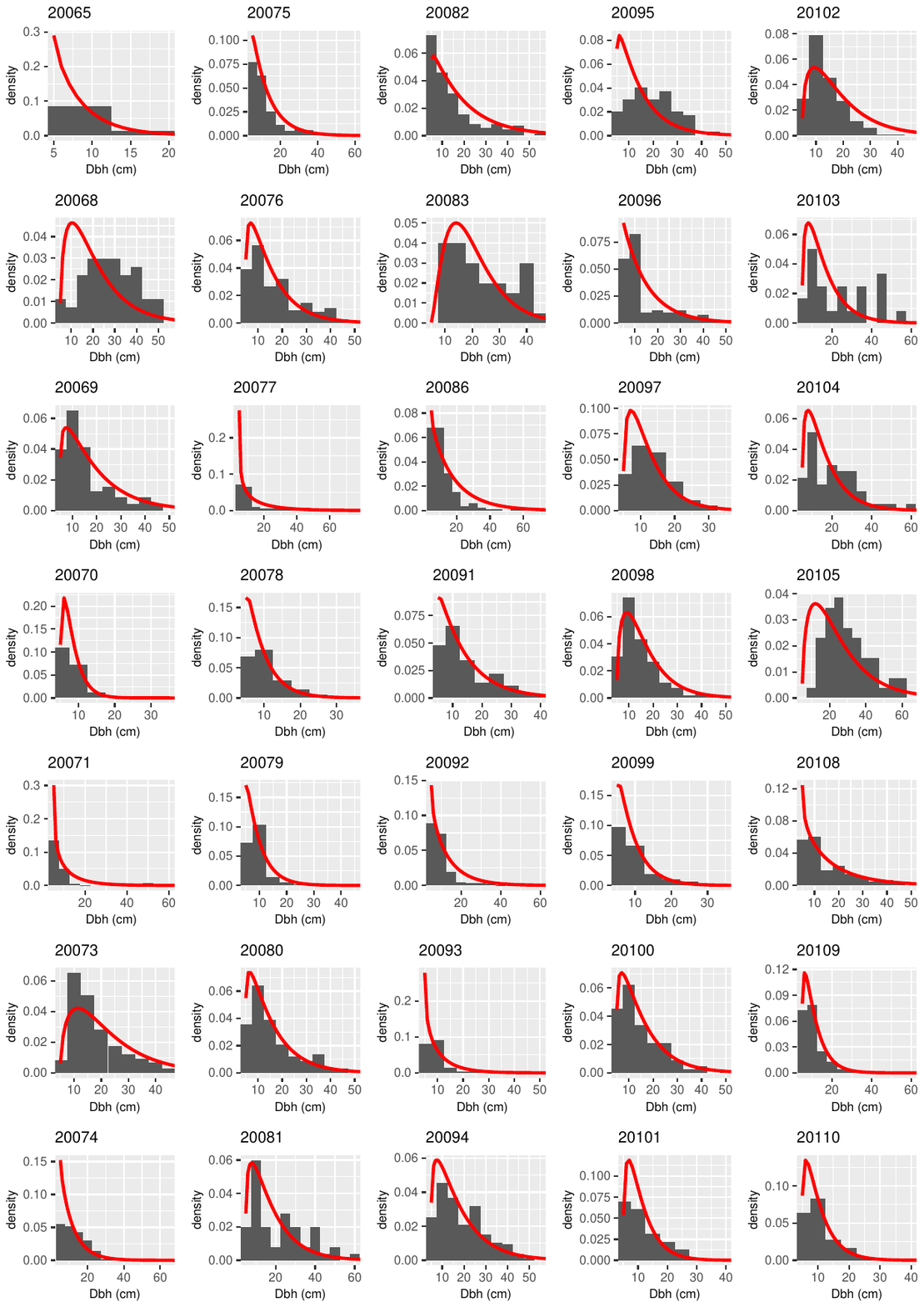}
\caption{Histograms and posterior predictive distributions of the DBH at the sample plots.}
\label{fig:dbh_pred_sampleplots_3}
\end{figure*}

\begin{figure*}
\centering
\includegraphics[height=1\textheight, page=4]{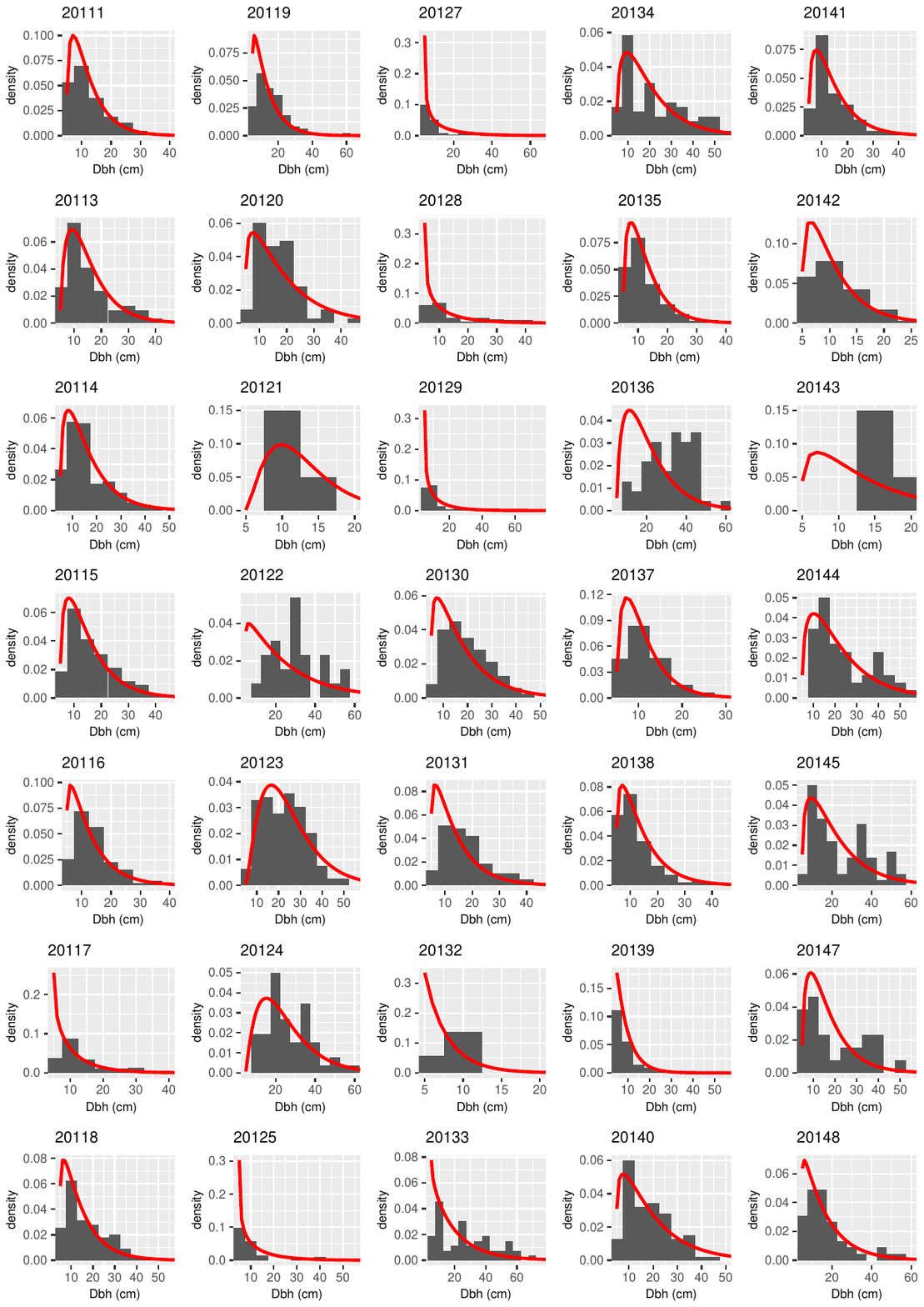}
\caption{Histograms and posterior predictive distributions of the DBH at the sample plots.}
\label{fig:dbh_pred_sampleplots_4}
\end{figure*}

\begin{figure*}
\centering
\includegraphics[height=1\textheight, page=5]{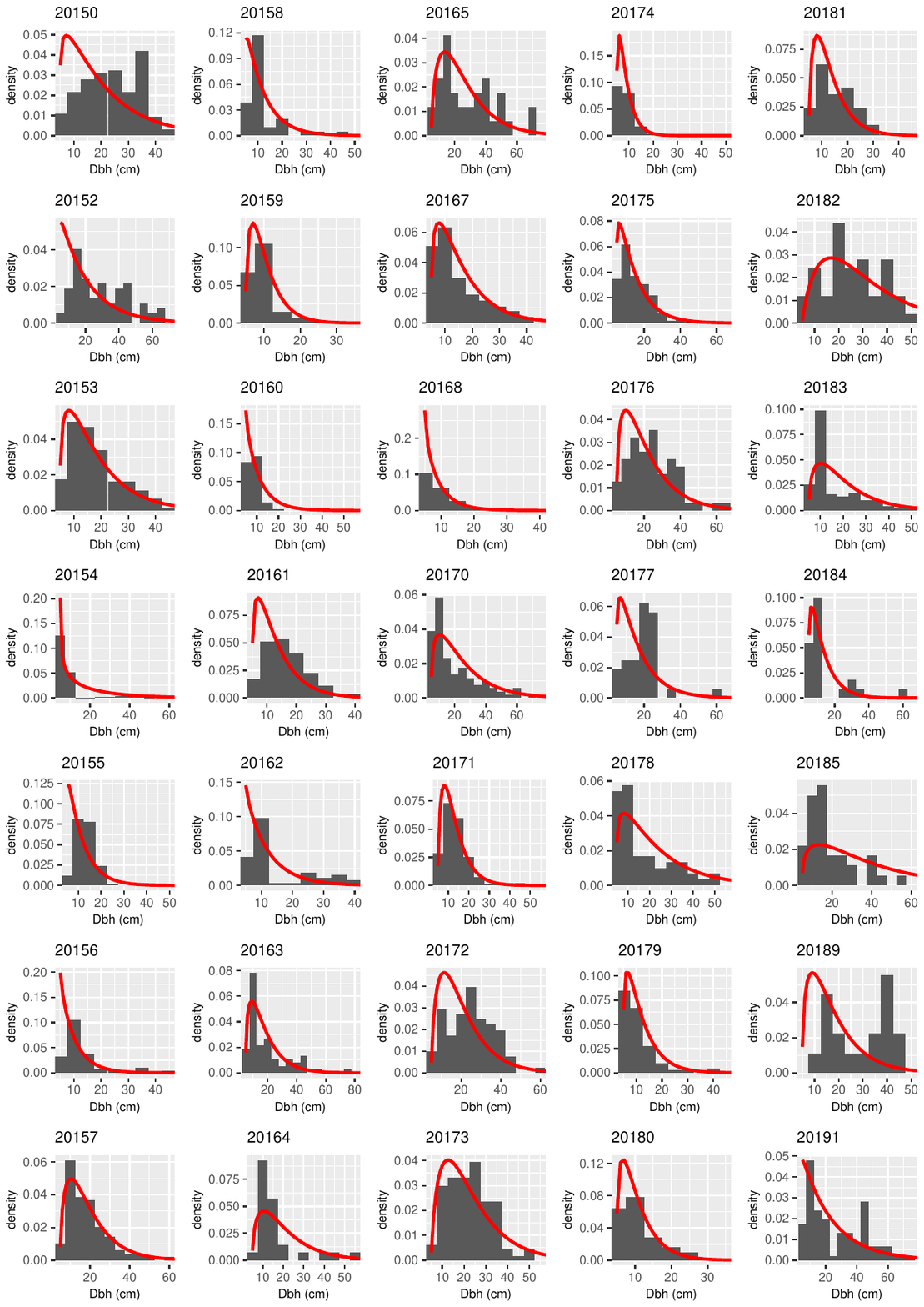}
\caption{Histograms and posterior predictive distributions of the DBH at the sample plots.}
\label{fig:dbh_pred_sampleplots_5}
\end{figure*}

\begin{figure*}
\centering
\includegraphics[height=1\textheight, page=6]{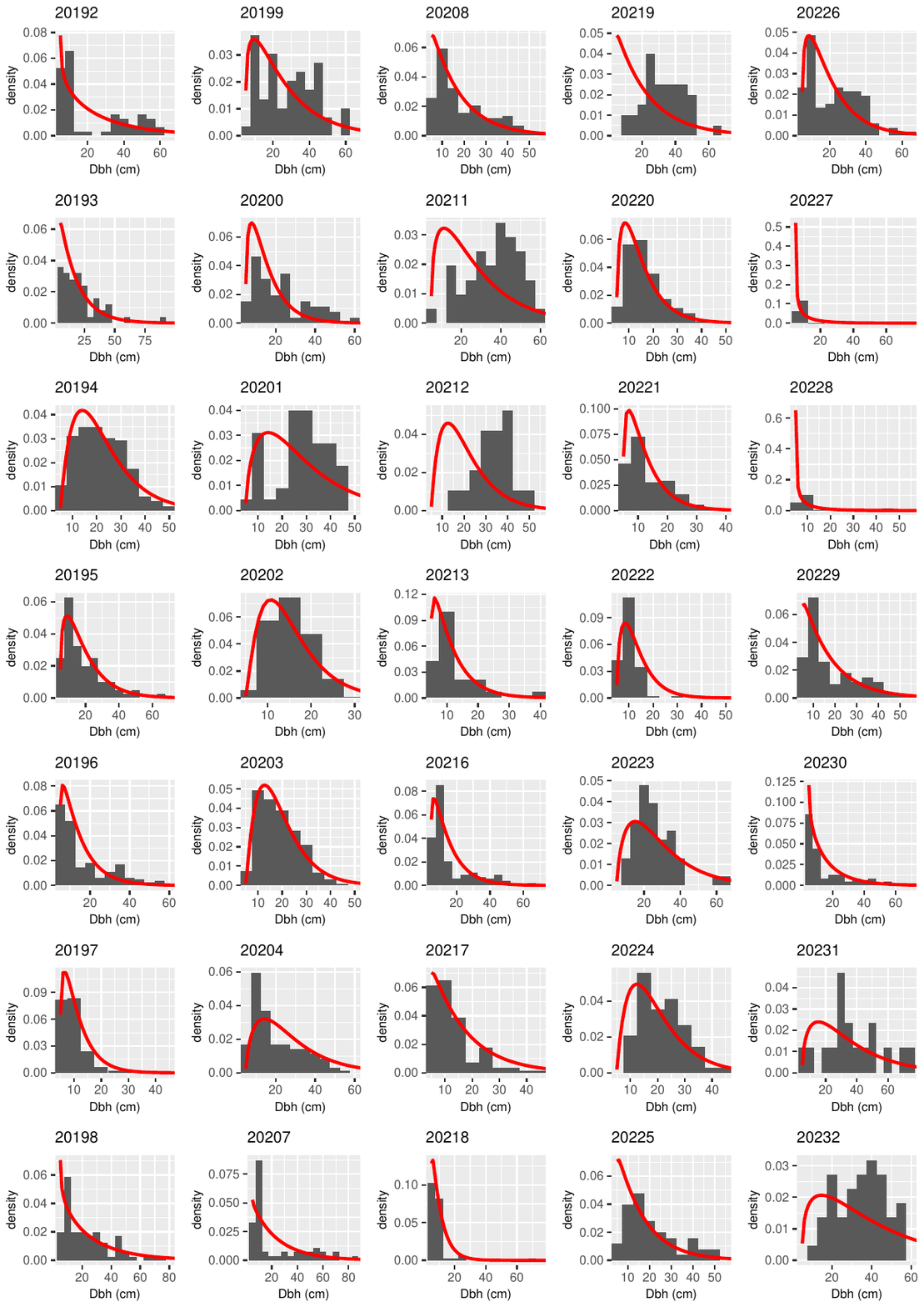}
\caption{Histograms and posterior predictive distributions of the DBH at the sample plots.}
\label{fig:dbh_pred_sampleplots_6}
\end{figure*}

\begin{figure*}
\centering
\includegraphics[height=1\textheight, page=7]{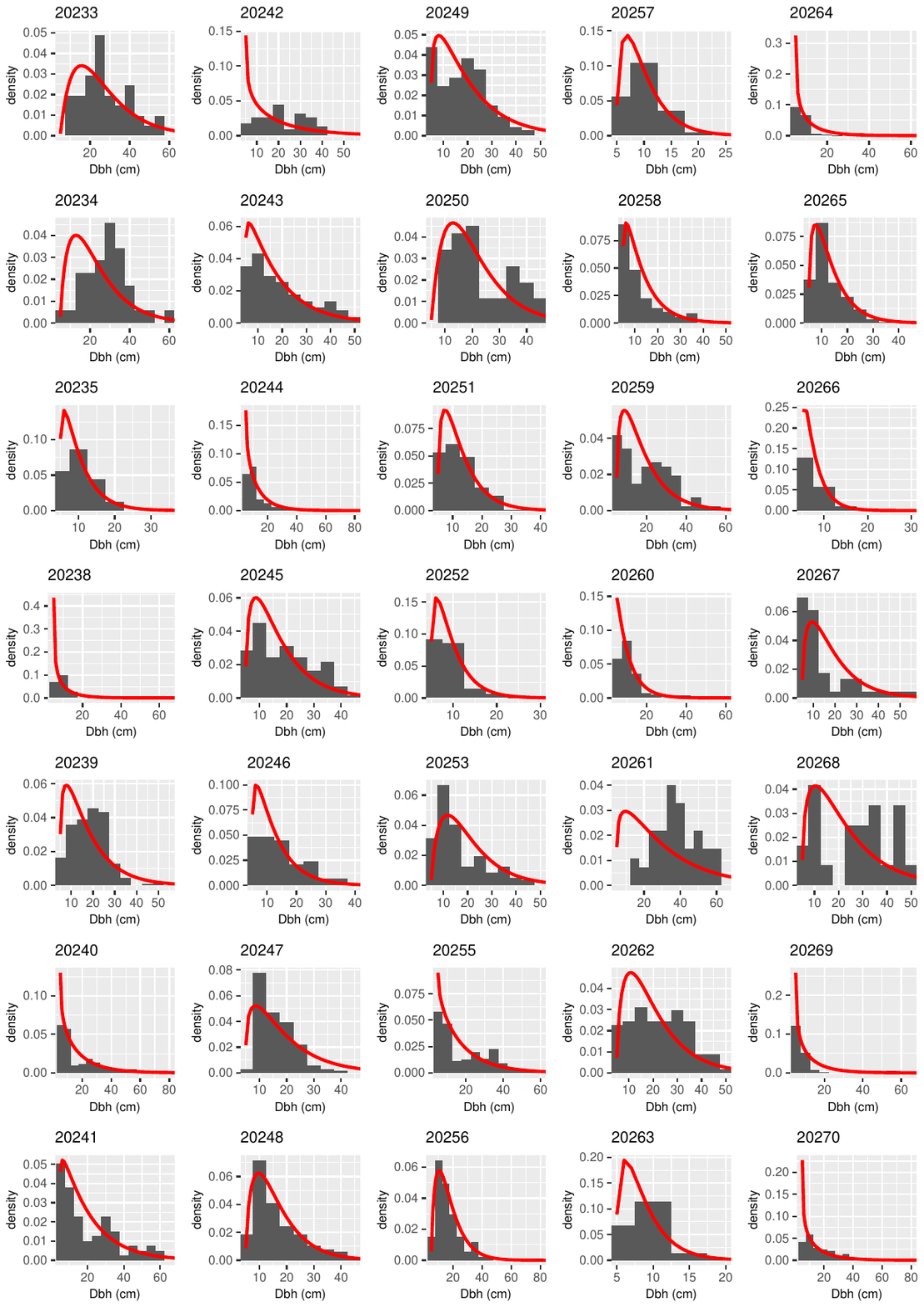}
\caption{Histograms and posterior predictive distributions of the DBH at the sample plots.}
\label{fig:dbh_pred_sampleplots_7}
\end{figure*}

\begin{figure*}
\centering
\includegraphics[height=1\textheight, page=8]{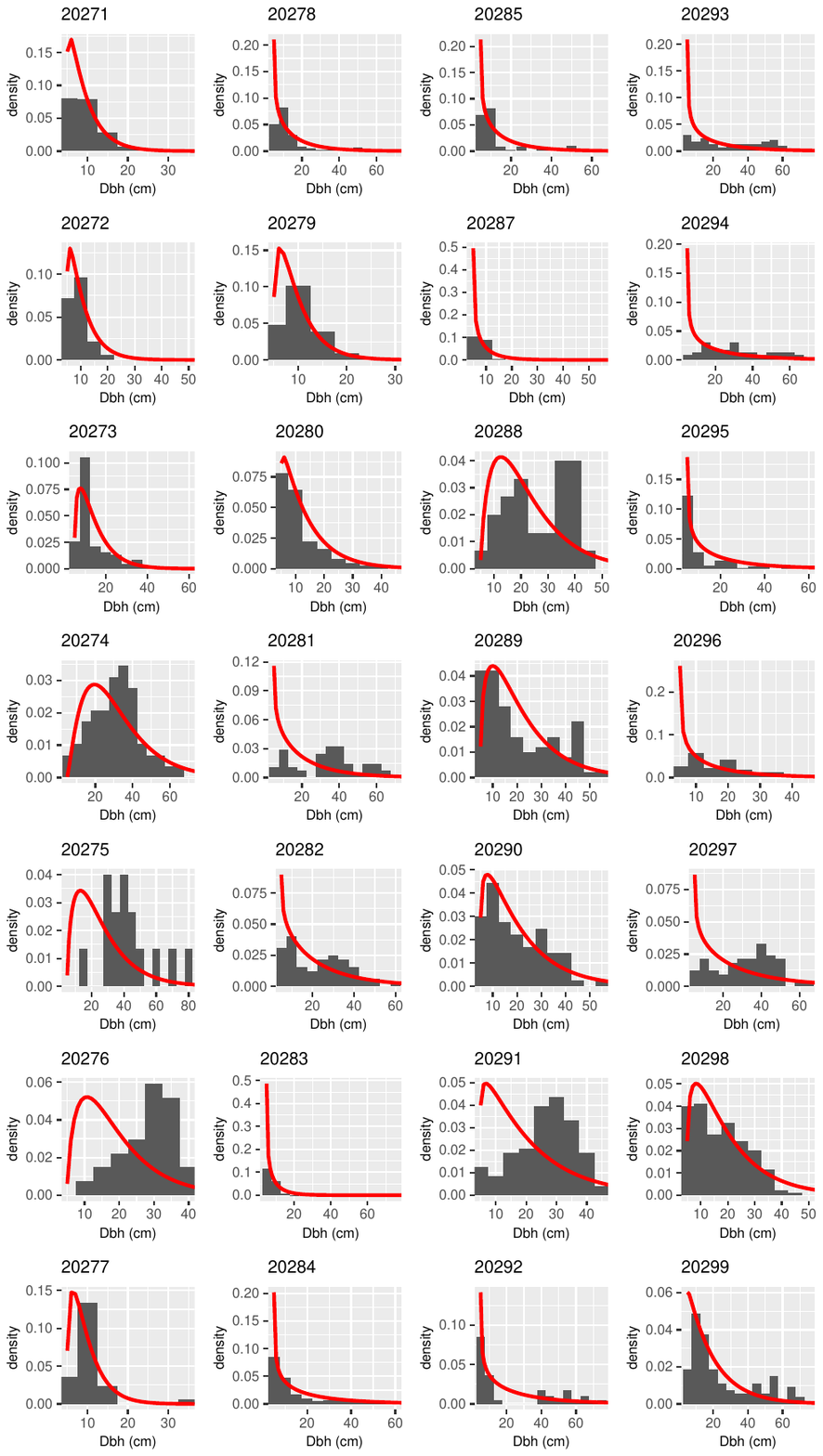}
\caption{Histograms and posterior predictive distributions of the DBH at the sample plots.}
\label{fig:dbh_pred_sampleplots_8}
\end{figure*}

\end{document}